\documentclass[proceedings]{JHEP} % 10pt is ignored!

\usepackage{epsfig}			% please use epsfig.
\newbox\mybox
\newcommand{\ttbs}{\char'134}           % \backslash for \tt (Nucl.Phys. :)%
\newcommand\fverb{\setbox\mybox=\hbox\bgroup\verb}
\newcommand\fverbdo{\egroup\medskip\noindent\fbox{\unhbox\mybox}\ }
\newcommand\fverbit{\egroup\item[\fbox{\unhbox\mybox}]}

\font\beeg=cmr17 scaled 1600		% Stylish initials
\newcommand\init[1]{\setbox\mybox=\hbox{{\beeg #1}~}%
		   \noindent\global\hangindent=\wd\mybox\global\hangafter-2%
		   \sc\smash{\llap {\lower 13.2pt \box\mybox}}}
\title{PERSPECTIVES IN HIGH-ENERGY PHYSICS}

\author{John Ellis
\\
	Theoretical Physics Division, CERN, CH-1211 Geneva 23\\
	E-mail: \email{John.Ellis@cern.ch}}

\conference{SILAFAE III, Cartagena de Indias, Colombia, April 2-8, 2000}

\abstract{A personal view of current prospectives in particle physics is
presented, inspired by the contributions to this meeting. Particular
emphasis is laid in precision tests of the Standard Model and the search
for the Higgs boson, on probes of CP violation, on speculations about
possible physics beyond the Standard Model, on neutrino masses and
oscillations, on the quest for supersymmetry, on opportunities @ future
accelerators, and on the ultimate phenomenological challenge offered by
the quest for a Theory of Everything.}

\dedicated{
\begin{center}
CERN-TH/2000-129 ~~~~~~~~~~~~ hep-ph/0007161\\
\end{center}
}

\begin{document} 

\maketitle %%%%%%%%%% THIS IS IGNORED %%%%%%%%%%%
%%%%%%%%%%%%%%%%%%%%%%%%%%%%%%%%%%%%%%%%%%%%%%%%%%%%%%%%%%%%%%%%%%%%%%%%%
\def\beq{\begin{equation}}
\def\eeq{\end{equation}}
\def\bea{\begin{eqnarray}}
\def\eea{\end{eqnarray}}
\def\bq{\begin{quote}}
\def\eq{\end{quote}}

\def\AJ{{\it Astrophys.J.} }
\def\AJL{{\it Ap.J.Lett.} }
\def\AJS{{\it Ap.J.Supp.} }
\def\AM{{\it Ann.Math.} }
\def\AP{{\it Ann.Phys.} }
\def\APJ{{\it Ap.J.} }
\def\APP{{\it Acta Phys.Pol.} }
\def\ASAS{{\it Astron. and Astrophys.} }
\def\BAMS{{\it Bull.Am.Math.Soc.} }
\def\CMJ{{\it Czech.Math.J.} }
\def\CMP{{\it Commun.Math.Phys.} }
\def\FP{{\it Fortschr.Physik} }
\def\HPA{{\it Helv.Phys.Acta} }
\def\IJMP{{\it Int.J.Mod.Phys.} }
\def\JMM{{\it J.Math.Mech.} }
\def\JP{{\it J.Phys.} }
\def\JCP{{\it J.Chem.Phys.} }
\def\LNC{{\it Lett. Nuovo Cimento} }
\def\SNC{{\it Suppl. Nuovo Cimento} }
\def\MPL{{\it Mod.Phys.Lett.} }
\def\NAT{{\it Nature} }
\def\NC{{\it Nuovo Cimento} }
\def\NP{{\it Nucl.Phys.} }
\def\PL{{\it Phys.Lett.} }
\def\PR{{\it Phys.Rev.} }
\def\PRL{{\it Phys.Rev.Lett.} }
\def\PRTS{{\it Physics Reports} }
\def\PS{{\it Physica Scripta} }
\def\PTP{{\it Progr.Theor.Phys.} }
\def\RMPA{{\it Rev.Math.Pure Appl.} }
\def\RNC{{\it Rivista del Nuovo Cimento} }
\def\SJPN{{\it Soviet J.Part.Nucl.} }
\def\SP{{\it Soviet.Phys.} }
\def\TMF{{\it Teor.Mat.Fiz.} }
\def\TMP{{\it Theor.Math.Phys.} }
\def\YF{{\it Yadernaya Fizika} }
\def\ZETF{{\it Zh.Eksp.Teor.Fiz.} }
\def\ZP{{\it Z.Phys.} }
\def\ZMP{{\it Z.Math.Phys.} }

\parskip 0.3cm

\def\gappeq{\mathrel{\rlap {\raise.5ex\hbox{$>$}}
{\lower.5ex\hbox{$\sim$}}}}

\def\lappeq{\mathrel{\rlap{\raise.5ex\hbox{$<$}}
{\lower.5ex\hbox{$\sim$}}}}

\section{The Standard Model}

\begin{figure}%WW
%\hglue4.5cm
\epsfig{figure=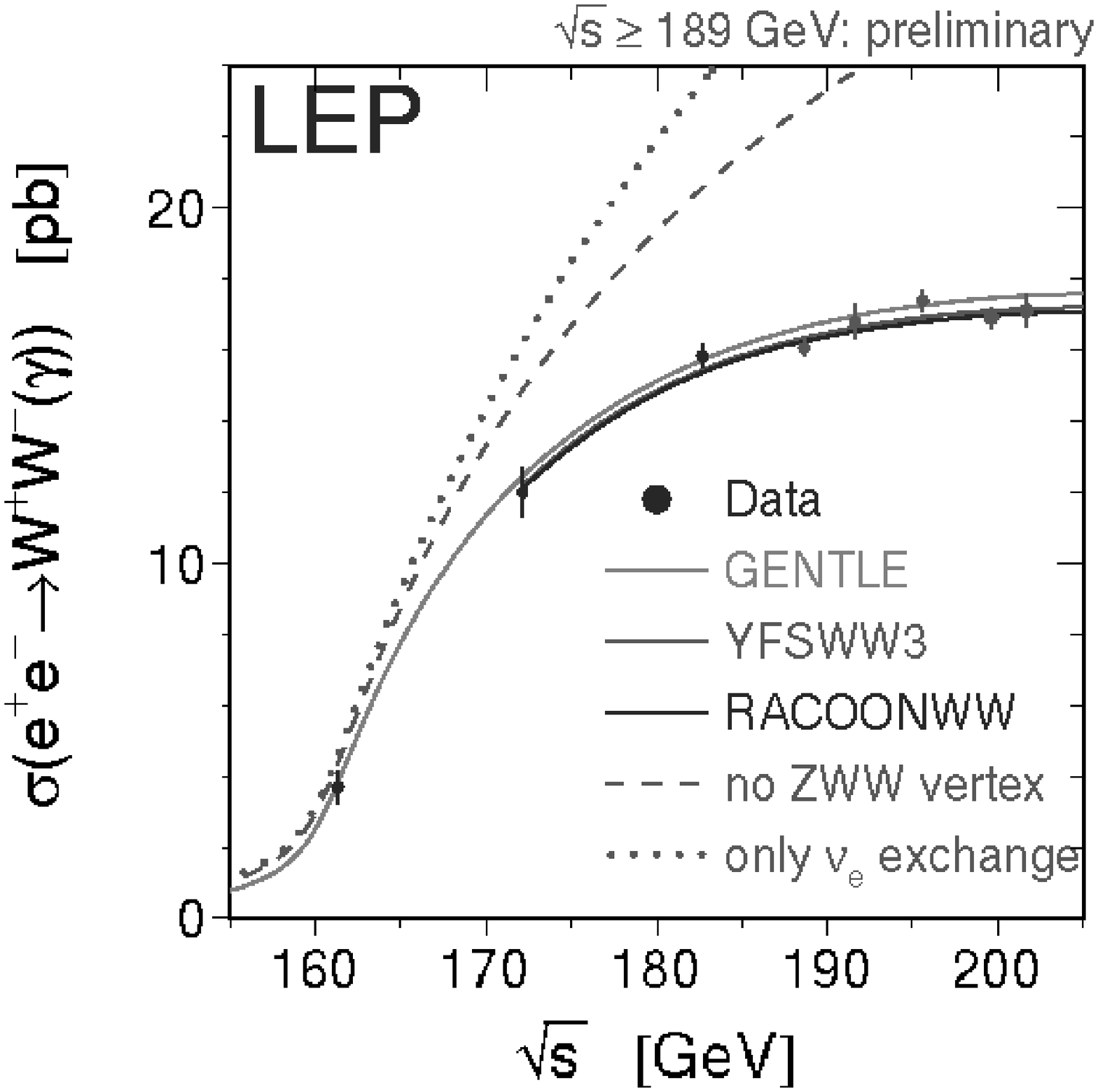,width=7cm}
\caption{\it The Standard Model rules OK: measurements of $\sigma (e^+
e^- \rightarrow W^+ W^-)$ at LEP~2 agree with theory, demonstrating the
existence of the expected $\gamma W^+ W^-$ and $ Z^0 W^+ W^-$
vertices~\cite{LEPEWWG}.}
\label{fig:WW}
\end{figure}

The Standard Model continued to rule accelerator experiments during 1999,
even as the heroic efforts of the CERN accelerator engineeers pushed the
LEP centre-of-mass energy to 202 GeV, and briefly to 204 GeV. There were
no surprises in fermion-pair production or in the bread-and-butter
reaction of LEP2, $e^+e^-\rightarrow W^+W^-$. Both the $\gamma W^+W^-$ and
$Z^0W^+W^-$ triple-gauge-boson vertices are there, as seen in
Fig.~\ref{fig:WW}, with magnitudes close
to the Standard Model values~\cite{LEPEWWG,Behner}. Looking into the final
states, there are no
confirmed interferences in $(W^+\rightarrow \bar qq)\otimes(W^-\rightarrow
\bar qq)$ final states, due to either colour rearrangement or
Bose-Einstein correlations: the difference between the $W$ mass measured
in purely hadronic and other final states is 15 $\pm$ 55
MeV~\cite{Nowell}. Combining
all the LEP measurements, one finds~\cite{LEPEWWG}

%\FIGURE[pos]{body}
%\smallskip
%\FIGURE{\epsfig{file=figWW.eps,width=7cm} 
        %\caption{\it The Standard Model rules OK: measurements of $\sigma (e^+
%e^- \rightarrow W^+ W^-)$ at LEP~2 agree with theory, demonstrating the
%existence of the expected $\gamma W^+ W^-$ and $ Z^0 W^+ W^-$
%vertices~\cite{LEPEWWG}.}
	%\label{WW}}

\beq
m_W = 80.401 \pm 0.048~{\rm GeV ~(LEP)},
\label{one}
\eeq
contributing with the hadron colliders ($M_W = 80.448 \pm 0.062$~GeV) to a
global average
\beq
m_W = 80.419 \pm 0.038~{\rm GeV ~(world)}.
\label{two}
\eeq
This error is now comparable with the value estimated indirectly from precision
electroweak measurements: $m_W = 80.382 \pm 0.026$~GeV, provides new,
independent evidence for a light Higgs
boson, and begins to impact significantly the radiative-correction
estimate~\cite{LEPEWWG}
\beq
m_H = 77^{+69}_{-39} \; {\rm GeV}
\label{three}
\eeq
when $\alpha_{em}(m_Z)^{-1} = 128.878 \pm 0.090$ is assumed (or
log$(M_H/{\rm GeV}) = 1.96^{+0.21}_{-0.23}$
if the estimate $128.905 \pm 0.036$, with the error reduced by
theory, is assumed). The Higgs boson probably weighs less than 200 GeV.

The plan is to raise the LEP energy as high as possible during 2000, with
the primary aim of searching for the Higgs boson. An integrated luminosity
of 50/pb per experiment at 206 GeV would increase the sensitivity of the
Higgs reach from the current lower limit of 107.9
GeV~\cite{LEPHiggs,Smith} to about 114 GeV~\cite{GR}.  A small numerical
increase, but in the most interesting range, also from the point of view
of supersymmetry~\cite{susyH}. The LEP energies attained so far range up
to 208.7~GeV, with a total luminosity (so far) of 109~pb$^{-1}$ at an
average
energy above 205~GeV. It seems that the target sensitivity to $m_H =
114$~GeV is well within reach. At the time of writing, the current
sensitivity is to $m_H \sim 113.4$~GeV, and the latest update may be
obtained from~\cite{Janot}. 

%\begin{figure}%GR
%\hglue4.5cm
%\hspace*{-2.0cm}
%\epsfig{figure=figGR.eps,width=12cm}
%\vspace*{-4.7cm}
%\caption{\it The sensitivity to $m_H$ obtainable from LEP running, as
%a function of the luminosity obtained at each energy~\cite{GR}. Also
%indicated is the sensitivity: $m_H \sim 109.2$~GeV expected from running 
%at 202~GeV during 1999.} 
%\label{fig:GR}
%\end{figure}

%\begin{figure}%dlds
%\hglue4.5cm
%\epsfig{figure=luday.eps,width=7cm}
%\caption{\it The luminosity delivered by LEP in various energy bands so
%far during 2000~\cite{luday}.}
%\label{fig:dlds}
%\end{figure}

\begin{figure}%Carena
%\hglue4.5cm
\epsfig{figure=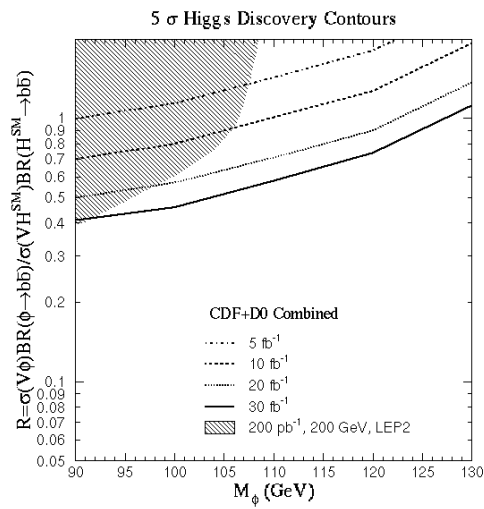,width=7cm}
\caption{\it Higgs discovery sensivity of the Fermilab Tevatron collider
compared with that of LEP~\cite{CMW}.}
\label{fig:Carena}
\end{figure}

Then, in Autumn 1999, LEP must be shut down and dismantled to make way for
the LHC excavations and installation. It will be the end of an era of
precision electroweak measurements. The search for the Higgs boson will
then pass to Fermilab, where the Tevatron has a chance of exploring higher
Higgs masses if it gathers more than 10 fb$^{-1}$ of luminosity, as seen
in Fig.~\ref{fig:Carena}~\cite{CMW}. 

\section{CP Violation}

For 24 years, it was possible to think that the CP violation~\cite{Nir} 
seen in the
$K^0-\bar K^0$ system might be due entirely to a superweak force beyond
the Standard Model, inducing CP violation in the $K^0-\bar K^0$ mass
matrix: Im$ M_{12} \not= 0$. Thanks to the new measurements of
$\epsilon^\prime/\epsilon$ in 1999~\cite{KTeV,NA48}, signalling direct CP
violation in the $K^0\rightarrow 2\pi$ decay amplitudes, we now know the
superweak theory cannot be the whole story. The Standard Model with 6
quarks permits CP violation in the $W^\pm$ couplings~\cite{KM}, but there
are no other sources if there is just one Higgs doublet. The MSSM contains
many possible sources of CP violation, of which the most important for
hadron decays may be the phases of the trilinear soft
supersymmetry-breaking couplings $A_{t,b}$ of the third generation and of
the gluino mass, relative to the Higgs mixing parameter $\mu$.

\begin{figure}%NA48
%\hglue4.5cm
\epsfig{figure=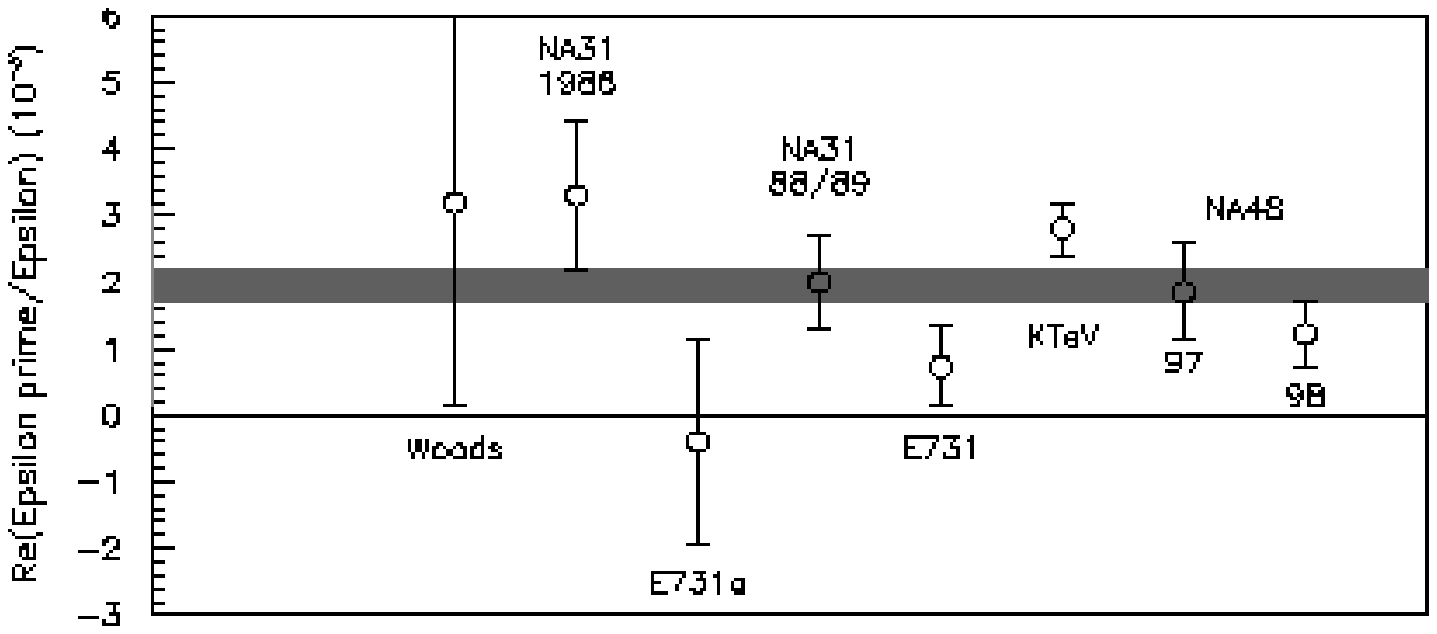,width=7cm}
\caption{\it Compilation of measurements of ${\rm Re}(\epsilon^\prime /
\epsilon )$~\cite{newNA48}.}
\label{fig:NA48}
\end{figure}

In the Standard Model, one may estimate~\cite{epsilonprime}
\begin{eqnarray}
& {\rm Re} \left( {\epsilon^\prime\over\epsilon}\right) 
\simeq 13 ~{\rm
Im}
\lambda_t
\left({130~{\rm MeV}\over
m_s(m_c)}\right)^2~\left({\Lambda^{(4)}_{\overline{MS}}\over 340~{\rm
MeV}}\right) \nonumber \\
& \times \left[ B_6 (1-\Omega) - 0.4 B_8
\left({m_t(m_t)\over 165~{\rm GeV}}\right)^{2.5}\right]
\label{four}
\end{eqnarray}
where Im$\lambda_t$ is a Cabibbo-Kobayashi-Maskawa angle factor and
$\Omega$ is an electroweak penguin effect. Calculating
Re$(\epsilon^\prime/\epsilon)$ accurately is difficult because of potential
cancellations in (\ref{four}), and the fact that the strong-interaction matrix
elements $B_6$ and $B_8$ are relatively poorly known. A conservative estimate in
the Standard Model would be that~\cite{epsilonprime}
\beq
1\times 10^{-4} < {\rm Re} (\epsilon^\prime / \epsilon ) < 3\times 10^{-3}
\label{five}
\eeq
The world data, including the new NA48 measurement reported
here~\cite{newNA48}, average to
\beq
{\rm Re}(\epsilon^\prime / \epsilon ) = (19.3 \pm 2.4)\times 10^{-4}
\label{six}
\eeq
as seen in Fig.~\ref{fig:NA48}.
This is compatible with the Standard Model range (\ref{five}), 
although probably
requiring a relatively big penguin matrix element $B_6$: an emperor maybe?
The measurement of $(\epsilon^\prime / \epsilon )$ will be refined by further
KTeV and NA48 data, and by KLEO at DAFNE. 

If the relatively large value of $(\epsilon^\prime / \epsilon )$ has a
large supersymmetric contribution, this could show up in $K^0_L\rightarrow
\pi^0\bar\nu\nu$, $\pi^0e^+e^-$ and $\pi^0\mu^+\mu^-$
decays~\cite{Masiero}. The present upper limits on these decays from KTeV
are far above the Standard Model predictions, so here is a physics
opportunity worth pursuing in parallel with $B$ experiments.

\begin{figure}%AL
%\hglue4.5cm
\epsfig{figure=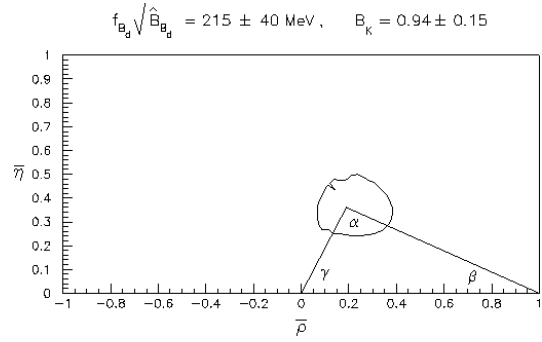,width=7cm}
\caption{\it The result of a recent global fit to the
Cabibbo-Kobayashi-Maskawa parameters~\cite{AL}.}
\label{fig:AL}
\end{figure}

The main target of these experiments will be the  Standard Model unitarity
triangle shown in Fig.~\ref{fig:AL}~\cite{AL}, with the hope of finding a
discrepancy: perhaps it
will turn out to be a quadrilateral? The poster child for the dawning CP age is
the measurement of $\sin 2\beta$ via $B^0\rightarrow J/\psi K_S$ decay. The
theory is gold-plated -- with penguin pollution expected only at the $10^{-3}$
level -- and the experiment is clean. Indeed, between them,
OPAL~\cite{OPALCP}, CDF~\cite{CDFCP} and ALEPH~\cite{ALEPHCP}
have almost measured it:
\beq
\sin 2 \beta = 0.91 \pm 0.35
\label{seven}
\eeq
A new era of precision flavour physics is now dawning, with
the $B$ factories now starting to take data. They should reduce the error
in $\sin 2 \beta$ below 0.1, and the LHC experiments aim at an error
$\sim$ 0.01.

That is the good news: the bad news concerns the measurement of $\sin 2\alpha$
via $B^0\rightarrow\pi^+\pi^-$ decay. This mode has now been seen by
CLEO~\cite{CLEO=hep-ex/0001010},
but with a relatively low branching ratio
\beq
B(\pi^+\pi^-) = (0.43^{+0.16}_{-0.14} \pm 0.05) \times 10^{-5}
\label{eight}
\eeq
whereas
\beq
B(K^+\pi^-) = (1.72^{+0.25}_{-0.24} \pm 0.12) \times 10^{-5}
\label{nine}
\eeq
This is bad news because it suggests that penguin pollution is severe, and
difficult to disentangle, as well as threatening a large background. This
worry has revived interest in the study of $\sin 2\alpha$ via
$B^0\rightarrow\rho^\pm\pi^\mp$ and other decays. There is no time here to
review proposed strategies for measured $\gamma$: suffice to say that
within a decade we may know the sum $\alpha + \beta + \gamma$ with a
precision $\sim 10^0$. Let us hope that it turns out to be inconsistent
with 180$^0$! 

\section{Beyond the Standard Model}

The Standard Model has many defects, the most severe being that it agrees
with all accelerator data. Nevertheless, it is very unsatisfactory, in
that it provides no explanations for the particle quantum numbers
$(Q_{em}, I, Y$, colour), and contains 19 arbitrary parameters: 3
independent gauge couplings, 1 CP-violating strong-interaction phase, 6
quark masses, 3 charged-lepton masses, 4 Cabibbo-Kobayashi-Maskawa mixing
parameters, the $W$ mass and the Higgs mass. 

As if this was not enough, the indications of neutrino oscillations introduce at
lest 9 more parameters: 3 neutrino masses, 3 neutrino mixing angles and 3
CP-violating phases, without even talking about the mechanism for neutrino mass
generation. Also, we should not forget gravity, with Newton's constant and the
cosmological constant as two new parameters, at least one more parameter
to generate the baryon
asymmetry of the Universe, at least one more to describe cosmological inflation,
etc. 

It is common to group the open problems beyond the Standard Model into
three categories: {\it Mass} -- do the particle masses indeed originate
from a Higgs boson, and is this accompanied by supersymmetric particles?
{\it Flavour} -- why are there so many flavours of quarks and leptons, and
what explains the ratios of their masses and weak mixings?  and {\it
Unification} -- is there a simple group structure containing the strong,
weak and electromagnetic interactions? Beyond these beyonds lurks the
quest for a {\it Theory of Everything} that should also include gravity,
reconcile it with quantum mechanics, explain the origin of space-time and
why it has 4 dimensions, and make Colombian coffee.  This is the ambition
of string theory and its latest incarnation as $M$ theory. 

Subsequent sections of this talk deal with these ideas and how they may be
tested. 

\section{Neutrino Masses and Oscillations}

Why not? Although the Standard Model predicts $m_\nu = 0$ if one ignores
possible non-renormaliz- able interactions, there is no
deep reason why this should be so. Theoretically, we expect masses to
vanish only if there is a good asymmetry reason, in the form of an exact
gauge symmetry. For example, the $U(1)$ gauge symmetry of QED guarantees
the conservation of electric charge and the masslessness of the photon. 
However, there is no exact gauge symmetry or massless gauge boson coupled
to lepton number $L$, which we therefore expect to be violated. Neutrino
masses are possible if there is an effective $\Delta L = 2$ interaction of
the Majorana form $m_\nu ~\nu \cdot \nu$. Such interactions are generic in
Grand Unified Theories (GUTs) with their extra particles, but could even
be fabricated from Standard Model particles alone~\cite{BEG}, if one
allows non-renormalizable interactions of the form ${1\over M} \nu H\cdot
\nu H \rightarrow m_\nu = <0\vert H\vert 0 >^2/M$, where $M$ is some heavy
mass scale:  $M \gg m_W$. 

Nevertheless, generic neutrino mass terms, as they arise in typical GUTs,
have the seesaw form~\cite{seesaw}
\beq
(\nu_L, \nu_R)~\left(\matrix{0&m \cr 
m^T&M}\right)~\left(\matrix{\nu_L\cr\nu_R}\right)
\label{ten}
\eeq
where the $\nu_R$ are singlet ``right-handed" neutrinos. After diagonalization,
(\ref{ten}) yields
\beq
m_\nu = m ~~{1\over M}~~ m^T
\label{eleven}
\eeq
which is naturally small: $m_\nu \ll m_{q,\ell}$, if $m\sim m_{q,\ell}$
and
$M =
{\cal O}(M_{GUT})$. For example, if $m\sim$ 10 GeV and $M\sim 10^{13}$ GeV, one
finds
$m_\nu \sim 10^{-2}$ eV, in the range indicated by experiments on solar and
atmospheric neutrinos. Each of the fields $(\nu_L,\nu_R)$ in (\ref{ten}) should
be regarded as a three-dimensional vector in generation space, so $m_\nu$
(\ref{eleven}) is a 3$\times$3 matrix. its diagonalization $V_\nu$ relative to
that of the charged leptons $V_\ell$ yields the Maki-Nakagawa-Sakata neutrino
mixing matrix $V_{MNS} = V_\ell V^+_\nu$ between the interaction eigenstates
$\nu_{e,\mu,\tau}$~\cite{MNS}.

Could there be additional light neutrinos? The LEP neutrino counting
measurements tell us that these could only be sterile neutrinos $\nu_s$.
But what would prevent them from acquiring large masses $m_s\nu_s\nu_s :
m_s \gg m_W$, since they have no electroweak quantum numbers to forbid
them via selection rules? This is exactly what happens to the $\nu_R$ in
(\ref{ten}). Most theorists expect the observed neutrinos $\nu_{light}\sim
\nu_L$, and their effective mass term to be of the Majorana type.
Before the advent of the atmospheric neutrino data, most theorists might
have favoured small neutrino mixing angles, by analogy with the CKM mixing
of quarks. But this is not necessarily the case, and many plausible models
have now been constructed in which the neutrino mixing is large, because
of large mixing in the light-lepton sector $V_\ell$ and/or in the heavy
Majorana mass matrix $M$ and hence $V_\nu$. 

In hierarchical
models of neutrino models, as often arise in GUTs: $m^2_1 \gg m^2_2 \gg
m^2_3$, one may interpret $\Delta m^2$ as $m^2_{\nu_H}$, the squared mass
of the
heavier eigenstate in the oscillation. Cosmological data exclude
$m_\nu \gappeq$ 3 eV, but even $m_\nu
\sim$
0.03 eV may be of cosmological importance~\cite{EHT}. Thus, although the
unconfirmed
LSND signal would certainly be of cosmological interest, so also 
are the
confirmed atmospheric neutrino data.

%\begin{figure}%Teg
%\hglue4.5cm
%\epsfig{figure=constraints.eps,width=7cm}
%\caption{\it Compilation~\cite{EHT} of constraints on neutrino masses and 
%mixing angles. The oscillation experiments provide indications on the
%mass-squared differences $\Delta m^2$, whereas cosmology is sensitive to
%the absolute mass values.} 
%\label{fig:Teg}
%\end{figure}

Even the solar neutrino data become of interest to cosmology if the three
neutrino flavous are almost degenerate, with $\Delta m^2 \ll m$, a possibility
that cannot be excluded by the oscillation data. Such a degeneracy is
constrained by the upper limit on neutrinoless $\beta\beta$
decay~\cite{Klapdor}:
\beq
< m_\nu >_e~~ \lappeq ~~ 0.2~{\rm eV}
\label{twelve}
\eeq 
where the expectation value is weighted by the neutrinos' electron
couplings. The limit (\ref{twelve}) is compatible with the neutrino masses
close to the cosmological upper limit (which nearly coincides with the
direct upper limit on $m_{\nu_e}$ from the end-point of Tritium $\beta$
decay) if there is almost maximal neutrino mixing. There is
an issue whether this can be maintained in the presence of the mass
renormalization expected in GUTs, which tend to break the mass degeneracy
and cause mixing to become non-maximal~\cite{EL}. However, these
difficulties may be avoided in some models of neutrino flavour
symmetries~\cite{Ross}. 

\begin{figure}%SK
%\hglue4.5cm
\epsfig{figure=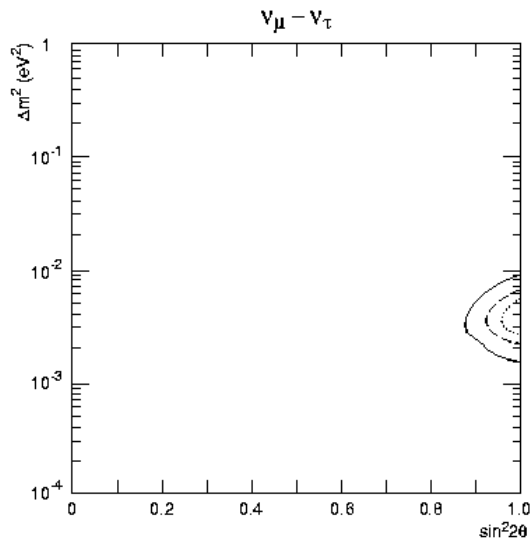,width=7cm}
\caption{\it Regions of neutrino oscillation parameter space favoured by
data on atmospheric neutrinos~\cite{SK}.}
\label{fig:SK}
\end{figure}

\begin{figure}%BKS
%\hglue4.5cm
\epsfig{figure=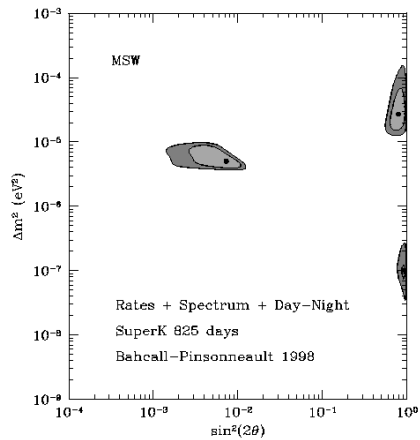,width=7cm}
\caption{\it MSW fits to the solar neutrino data~\cite{BKS}.}
\label{fig:BKS}
\end{figure}

As was discussed here by Smirnov~\cite{Smirnov} and is seen in
Fig.~\ref{fig:SK}, the best fit to the
Super-Kamiokande atmospheric-neutrino data is with near-maximal mixing and
$\Delta m^2 \simeq 3.7\times 10^{-3}$ eV, the 90 \% confidence-level range
being (2 to 7)$\times 10^{-3}$ eV$^2$~\cite{SK}. This is compatible with
the 90 \% confidence-level ranges favoured by the other
atmospheric-neutrino experiments (Kamiokande, MACRO and Soudan). As for
the solar-neutrino data, there have been four favoured regions, three
of which shown in
Fig.~\ref{fig:BKS}: the large- and small-mixing-angle MSW solutions (LMA
and SMA) with $\Delta m^2 \sim 10^{-5}$ eV$^2$, another MSW solution with
lower $\Delta m^2 \sim 10^{-7}$ eV$^2$ (LOW), and vacuum solutions (VAC)
with $\Delta m^2 \sim 10^{-9}$ to $10^{-11}$ eV$^2$~\cite{BKS}. 

How may one discriminate between the different oscillation scenarios? We
know already from Super-Kamiokande that $\nu_\mu\rightarrow\nu_e$
atmospheric oscillations are not dominant, and more stringent upper limits
are imposed by the Chooz and Palo Verde reactor experiments~\cite{Chooz}.
However,
$\nu_\mu\rightarrow\nu_e$ oscillations may be present at a lower level: if
so, they may enable the sign of $\Delta m^2$ to be determined via matter
effects. There are several ways to distinguish between dominance by
$\nu_\mu\rightarrow\nu_\tau$ or $\nu_\mu\rightarrow\nu_{s}$. 
The zenith-angle distributions in various categories of
Super-Kamiolande events (high-energy partially-contained events,
throughgoing muons and neutral-current enriched events) favour
$\nu_\mu\rightarrow\nu_\tau$ oscillations, now at the 99 \%
confidence level~\cite{Shiozawa}. Another way to distinguish
$\nu_\mu\rightarrow\nu_\tau$ from $\nu_\mu\rightarrow\nu_{s}$ is
$\pi^0$ production in Super-Kamiokande, which is impossible in $\nu_{s}$
interactions. The present data appear to favour
$\nu_\mu\rightarrow\nu_\tau$, but, before reaching a conclusion, 
it will be necessary to reduce the systematic
error in the production rate, which should be possible using data from the
K2K near detector. 

As for solar neutrinos, no significant distortion in the energy spectrum
is now observed by Super-Kamiokande~\cite{Suzuki}, and there is an upper
limit on the flux of hep
neutrinos that excludes the possibility that they might influence
oscillation interpretations. Moreover, the
day-night effect is now apparent only at the
1.3-$\sigma$ level, and the seasonal
variation that seems to be emerging is completely consistent with
the expected geometric effect of the Earth's orbital eccentricity.
These observations are all consistent with the LMA interpretation if
$\Delta m^2 \gappeq 2 \times 10^{-3}$~eV$^2$, but the constraints
from the day and night spectra
now disfavour the SMA and VAC interpretations at the 95 \%
confidence level~\cite{Suzuki}. Moreover, $\nu_e\rightarrow\nu_{s}$ is
also
disfavoured at the same level. Super-Kamiokande seems to be pushing
us towards the LMA $\nu_\mu\rightarrow\nu_\tau$ interpretation, but
has not yet provided a `smoking gun' for this interpretation.

In the near future, SNO will be providing information on the
neutral-current/charged-curr- ent ratio, discriminating between the
different solar-neutrino scenarios and telling us whether the $\nu_e$ have
oscillated into $\nu_\mu / \nu_\tau$ or $\nu_{s}$~\cite{SNO}. In the
longer term, BOREXINO will check the disappearance of the $^7$Be solar
neutrinos~\cite{BOREXINO}.

Many of the most important prospective developments may be provided by
long-baseline terrestrial neutrino experiments. The first of these is K2K,
which has already released some preliminary
results~\cite{K2K=hep-ex/0004015}.  They see 17 fully-contained events
in their fiducial region, whereas 29.2$^{+3.5}_{-3.3}$ would have been
expected in the absence of oscillations, 
versus 19.3$^{+2.5}_{-2.4}$ if $\delta m^2 = 3 \times 10^{-3}$.
Overall, they see a total
of 44 events, whereas about 74 (50) would have been expected in all event
categories in the absence of oscillations (if $\delta m^2 = 3 \times 10^{-3}$),
A detailed fit including data from the run currently underway and
energy-spectrum information is now being prepared, but the present data 
already appear to be very promising!
In a few years' time, KamLAND will use
reactor neutrinos to check the LMA MSW solution to the solar-neutrino
deficit~\cite{KamLAND}. Starting probably in 2003, MINOS will be looking
for $\nu_\mu$ disappearance, measuring the neutral-current/charged-current
ratio, and looking for $\nu_e$ appearance~\cite{MINOS}.

\begin{figure}%OPERA
%\hglue4.5cm
\epsfig{figure=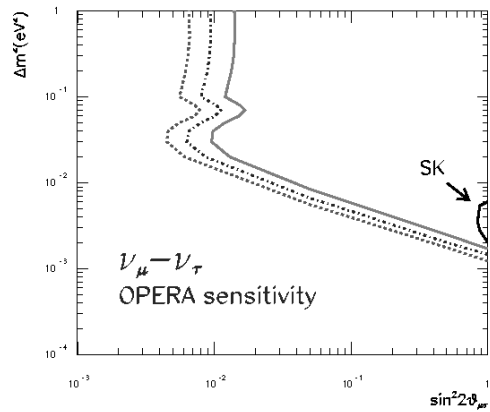,width=7cm}
\caption{\it Sensitivity to $\nu_\mu\rightarrow\nu_\tau$ oscillations
of the OPERA experiment~\cite{OPERA} proposed for the
CERN-Gran Sasso neutrino beam~\cite{CNGS}.}
\label{fig:OPERA}
\end{figure}

Then, starting in 2005, the CERN-Gran Sasso project will provide the
opportunity to look for $\nu_\mu\rightarrow\nu_\tau$ oscillations directly
via $\tau$ production~\cite{CNGS}. This seems to me a key experiment: ``If
you have not seen the body, you have not proven the crime." The beam
energy has been optimized for $\tau$ production, and either of the
proposed experiments (OPERA~\cite{OPERA} and ICANOE~\cite{ICANOE})  should
be able to discover $\tau$
production at the 4-$\sigma$ level if the atmospheric-neutrino parameters
are in the range favoured by Super-Kamiokande, as exemplified in
Fig.~\ref{fig:OPERA}. Additionally, ICANOE~\cite{ICANOE} may be able to
probe the LMA
MSW solar solution via low-energy atmospheric-neutrino events.

\section{Supersymmetry}

As you know, the motivation for supersymmetry at accessible energies is the
hierarchy problem~\cite{hierarchy}, namely why $m_W \ll m_P$ , or
equivalently why $G_F \gg G_N$,
or equivalently why the Coulomb potential dominates over the Newton potential in
atoms. Even if the hierarchy is set by hand, an issue is raised by the quantum
corrections:
\beq
\delta m^2_{H,W} = {\cal O} \left({\alpha\over\pi}\right) \Lambda^2 \gg m^2_W
\label{thirteen}
\eeq
where $\Lambda$ is a cutoff reflecting the appearance of new physics beyond the
Standard Model. Supersymmetry is a favoured example, making the corrections
(\ref{thirteen}) naturally small:
\beq
\delta m^2_{H,W} = {\cal O} \left({\alpha\over\pi}\right) ~~\left\vert m^2_B -
m^2_F\right\vert
\label{fourteen}
\eeq
which is $\lappeq m^2_{H,W}$ if
\beq
\vert m^2_B - m^2_F\vert \lappeq 1~{\rm TeV}^2
\label{fifteen}
\eeq
for the difference in mass-squared between spartners. Circumstantial
evidence for supersymmetry is provided by the concordance between the
gauge coupling strengths measured at LEP and elsewhere with supersymmetric
GUTs~\cite{g1g2g3}, and the indirect LEP indications for a light Higgs
boson~\cite{LEPEWWG}, as predicted by supersymmetry~\cite{susyH} and
discussed shortly. However, neither these nor the naturalness/fine-tuning
argument set rigorous upper bounds on the sparticle mass
scale~\cite{finetune}.

Another argument favouring low sparticle masses is provided by cold dark
matter~\cite{EHNOS}.
The lightest sparticle is commonly expected to be the lightest eigenstate of the
neutralino mass matrix:
\beq
\left(
\matrix{ M_2 & 0 & {-g_2v_2\over \sqrt{2}} & {g_2v_1\over\sqrt{2}} \cr
&&&\cr
0 & M_1 & {-g^\prime v_2\over \sqrt{2}} & {g^\prime v_1\over\sqrt{2}} \cr
&&&\cr
{-g_2v_2\over \sqrt{2}} & {-g^\prime v_1\over\sqrt{2}} &0 & \mu \cr
&&&\cr
{-g_2v_1\over \sqrt{2}} & {-g^\prime v_1\over\sqrt{2}} &\mu & 0 \cr}
\right)
\label{sixteen}
\eeq
The gaugino masses $M_2, M_1$ are commonly assumed to be equal at the GUT scale:
\beq
M_2 = M_1 \equiv m_{1/2}
\label{seventeen}
\eeq
and are then renormalized: $M_2/M_1 \simeq \alpha_2/\alpha_1$ at lower
scales.  The scalar masses may also be universal at the GUT scale, in
which case they are also renormalized: $m^2_{0_i} = m^2_0 + C_i m_{1/2}$,
and ratio of Higgs v.e.v.'s is commonly denoted by $\tan\beta \equiv
v_2/v_1$. 

\begin{figure}[htb]
\epsfig{file=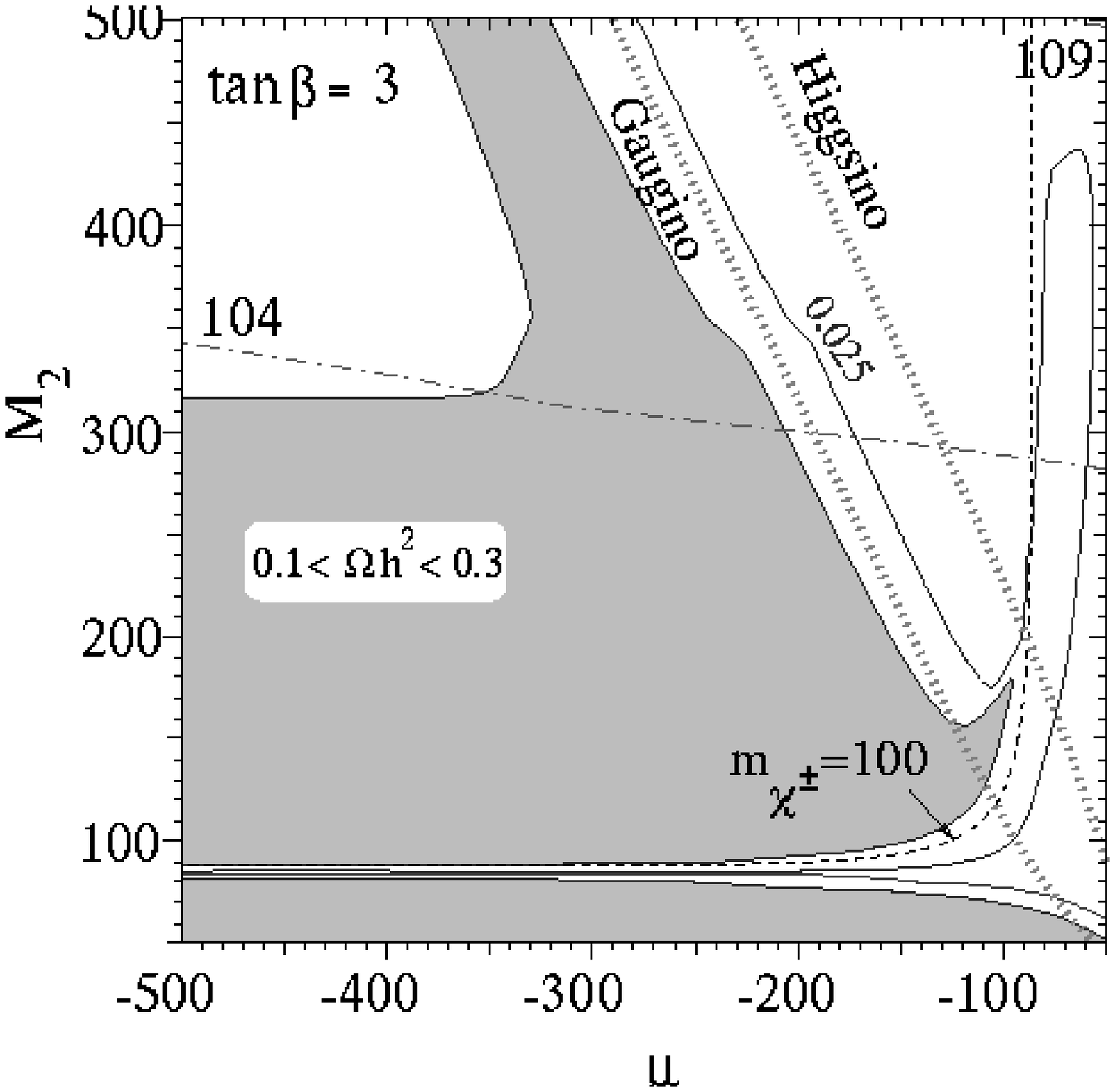,width=7.0cm}
\epsfig{file=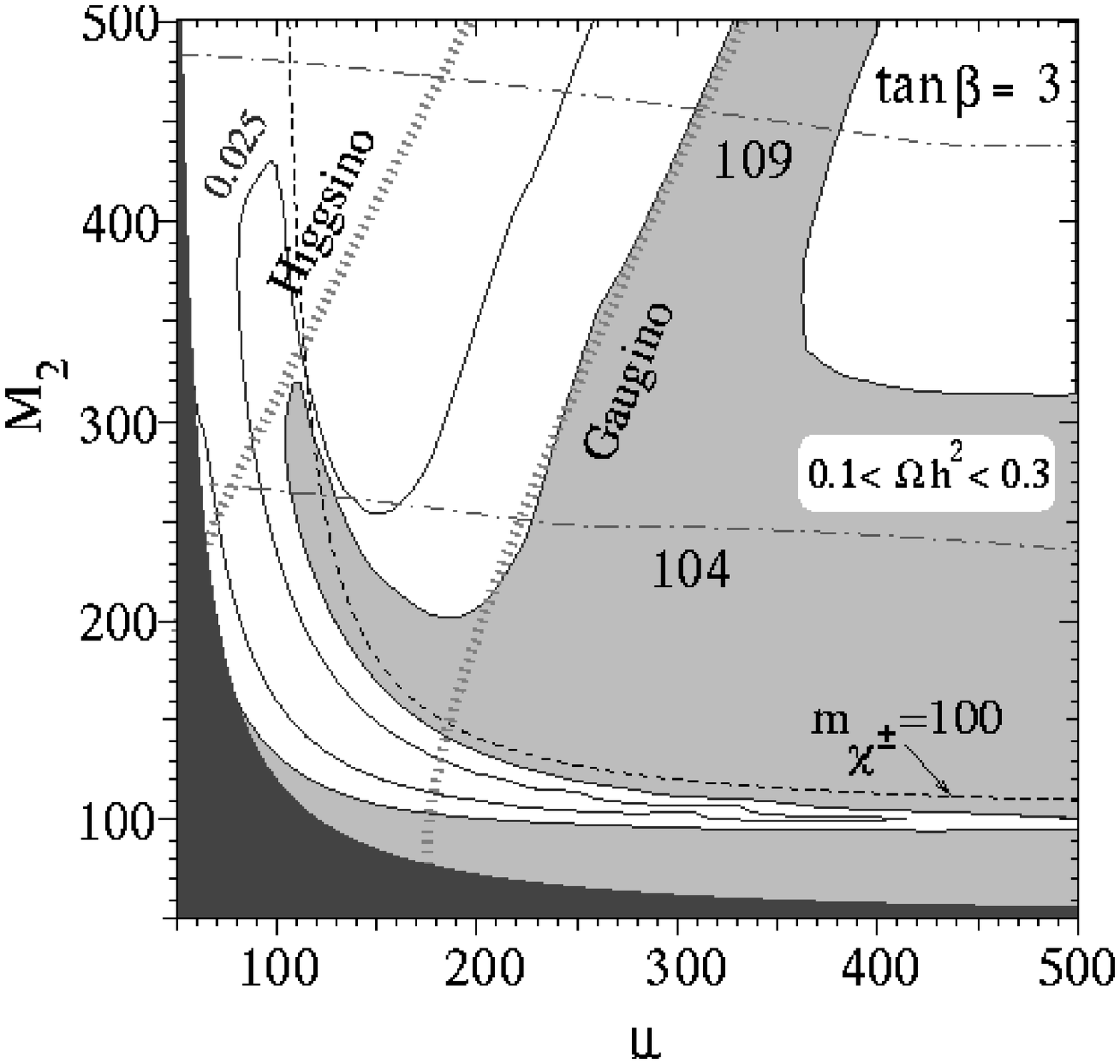,width=7.0cm}
\caption{\it Regions of the $\mu, M_2$ plane where the lightest
neutralino $\chi$ may constitute the cold dark matter~\cite{EFOS} (light
shading). In
most of the allowed regions, the $\chi$ is mainly a $U(1)$ gaugino. The
dark shaded region was excluded by LEP~1.} 
\label{fig:bino}
\end{figure}

The LEP limits on neutralinos and charginos~\cite{Fouchez} exclude the
possibility that the lightest
neutralino $\chi$ is an almost pure photino or Higgsino. If univerality is
assumed, or if neutralinos constitute the bulk of the cold dark
matter~\cite{Roulet}, a
dominant $U(1)$ gaugino component is favoured, as shown in
Fig.~\ref{fig:bino}~\cite{EFGO}.  
%Fig.~\ref{fig:m0mhalf} shows (for $\mu > 0$)
As discussed in~\cite{EFGO}, one must take into account 
the various constraints on the universal parameters $(m_{1/2}, m_0)$
imposed by LEP, the absence of charged $\tilde\tau$ dark matter, the
observed rate for $b\rightarrow s\gamma$ decay and the absence of a
charge- and colour-breaking (CCB) vacuum. The allowed dark-matter region
is stretched to large $m_{1/2}$ by neutralino-slepton
coannihilation, which allow $m_\chi \lappeq$ 600 GeV~\cite{EFOS}. 

%\begin{figure}%bino
%\hglue4.5cm
%\epsfig{figure=fig1n.eps,width=8cm}
%\hglue2.5cm
%\epsfig{figure=fig1p.eps,width=8cm}

%\begin{center}
%\mbox{\epsfig{file=fig1n.eps,height=7.5cm}}
%\mbox{\epsfig{file=fig1p.eps,height=7.5cm}}
%\end{center}

%\begin{figure}%bino
%\hglue4.5cm
%\epsfig{figure=fig1p.eps,width=8cm}
%\caption{}
%\end{figure}

The most direct experimental constraints on the supersymmetric parameter
space are those for charginos, neutralinos and sleptons at LEP, which
impose $m_{\chi^\pm} \gappeq$ 100 GeV and $m_{\tilde e} \gappeq$ 90 GeV
generically. The Tevatron constraints on squarks and gluinos are of less
direct importance if universality is assumed. However, stop searches at
LEP and the Tevatron are important for constraining the radiative
corrections to the lightest supersymmetric Higgs mass, whose leading terms
contribute~\cite{susyH}
\beq
\delta m^2_h = {\cal O} (\alpha)~~{m^4_t\over m^2_W}~~ \ln \left({m^2_{\tilde
t}\over m^2_t}\right)
\label{eighteen}
\eeq
These are relevant to the constraints on $m_0$ and $m_{1/2}$ imposed by the
absence of a Higgs boson at LEP~\cite{EFOS}.
%, also shown in
%Fig.~\ref{fig:m0mhalf}~\cite{EFOS}.

%\begin{figure}[htb]
%\vspace*{2.9 cm}
%\hspace*{-2.5 cm}
%\epsfig{file=fig9.eps,width=6.0cm}
%\epsfig{file=us3p.eps,height=3.5cm}
%\epsfig{file=us5p.eps,height=3.5cm}
%\epsfig{file=us10p.eps,height=3.5cm}
%\epsfig{file=us20p.eps,height=3.5cm}

%\caption{\it Regions of the $m_{1/2}, m_0$ planes where the lightest
%neutralino $\chi$ may constitute the cold dark matter, if one 
%assumes universal soft supersymmetry-breaking scalar masses
%$m_0$~\cite{EFOS,EFGO} (medium shading). The dotted (dashed)
%(dash-dotted)
%(long-dashed) lines
%are contours of $m_{\tilde e}$ ($m_{\chi^\pm}$) ($m_h$) (Tevatron Run~II
%sensitivity), and there are CCB minima below the solid lines. The dark
%shaded regions at the bottoms of the panels would have charged dark
%matter, and hence are disallowed, and the light shaded
%region on the left of panel (d) is
%excluded by measurements of $b \rightarrow s \gamma$.}
%\label{fig:m0mhalf}
%\end{figure}

\begin{figure}[htbp]
\epsfig{file=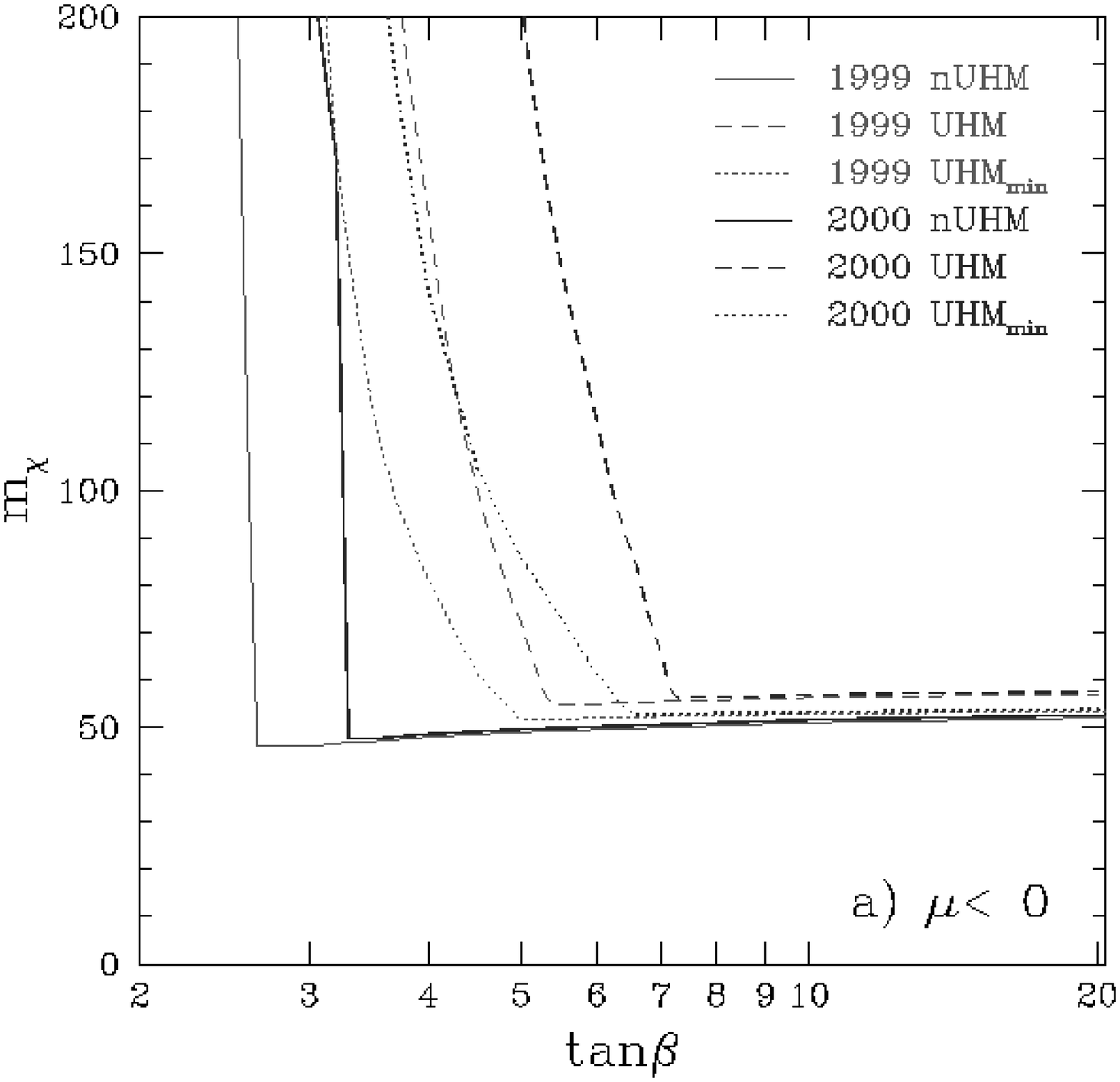,height=7.0cm}
\epsfig{file=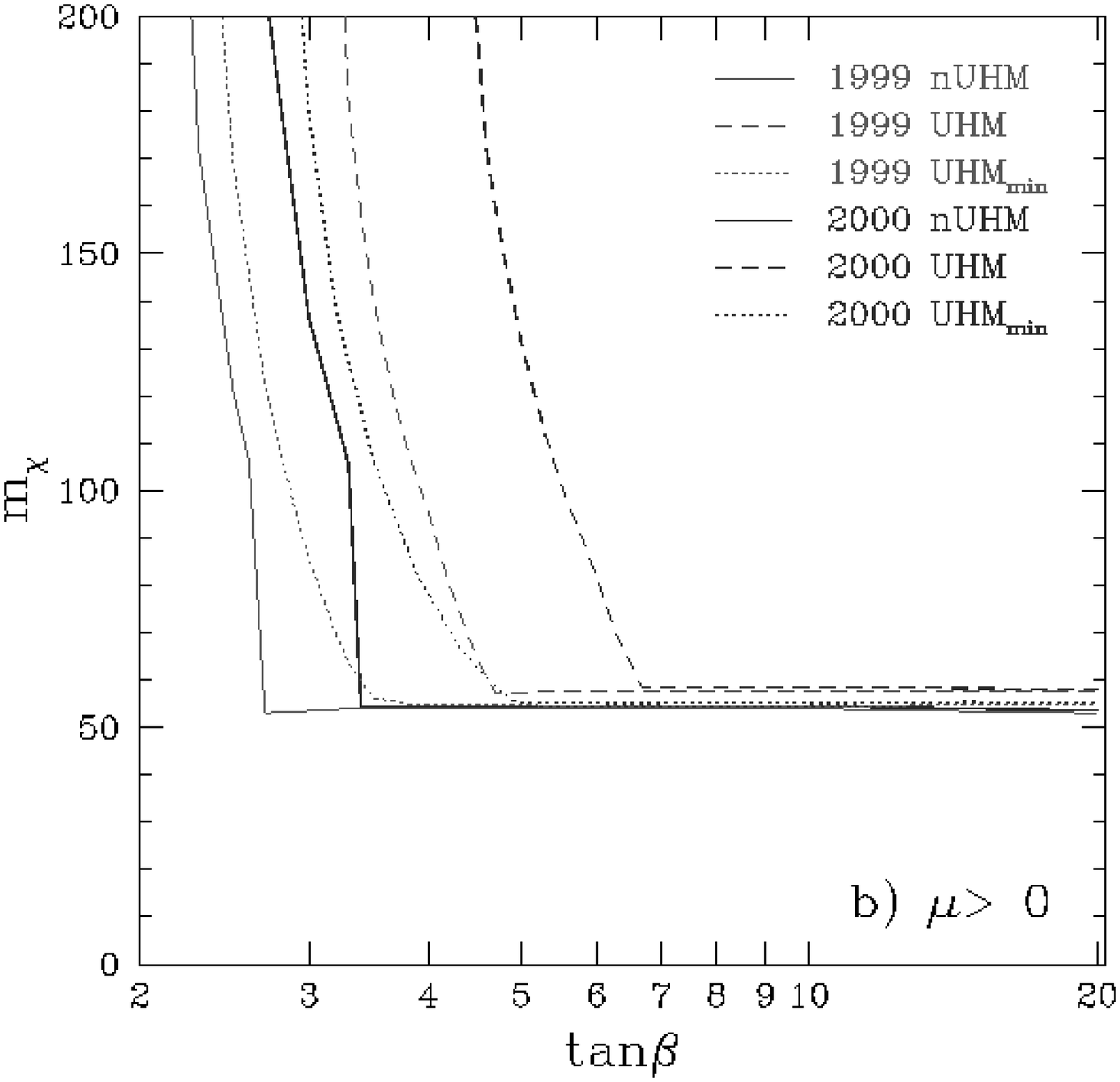,height=7.0cm}
\caption{\it Lower limits on the mass of the lightest neutralino $\chi$
obtained under various assumptions: universal scalar masses, also for
Higgs multiplets, and the absence of CCB minima (UHM) or allowing them
(UHM$_{\rm min}$), and non-universal scalar masses (nUHM)~\cite{EFGO}.}
{\label{fig:mchi}} 
\end{figure}

Fig.~\ref{fig:mchi} displays the lower limits on $m_\chi$ imposed by all
these
constraints, either if scalar-mass universality is (UHM) or is not (nUHM)
assumed~\cite{EFOS}. In the UHM$_{min}$ scenario in Fig.~\ref{fig:mchi},
the absence
of CCB vacua is not
required. Fig.~\ref{fig:mchi} also displays the expected impact of LEP
searches in 2000,
assuming pessimistically that they are unsuccessful. We see that
\beq
m_\chi \gappeq 50~{\rm GeV \quad\quad and} \quad\quad \tan\beta \gappeq 3~,
\label{nineteen}
\eeq
with the precise values depending on the scenario adopted.

\begin{figure}%CQW
%\hglue4.5cm
\epsfig{figure=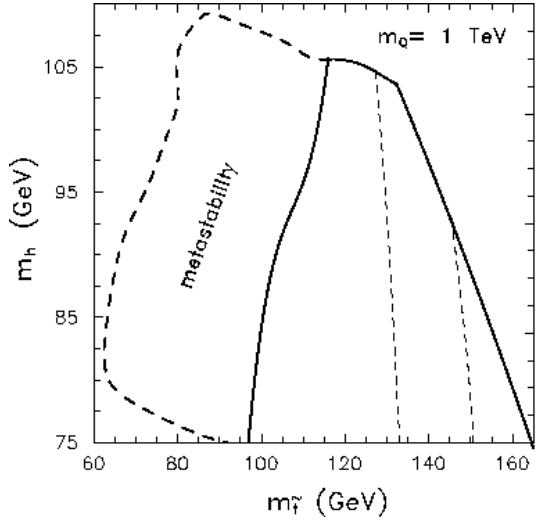,width=7cm}
\caption{\it Region of the $m_{\tilde t}, m_h$ plane where the
electroweak phase transition may be strongly first order~\cite{CQW}.}
\label{fig:CQW}
\end{figure}

\begin{figure}%HZZ
%\hglue4.5cm
\epsfig{figure=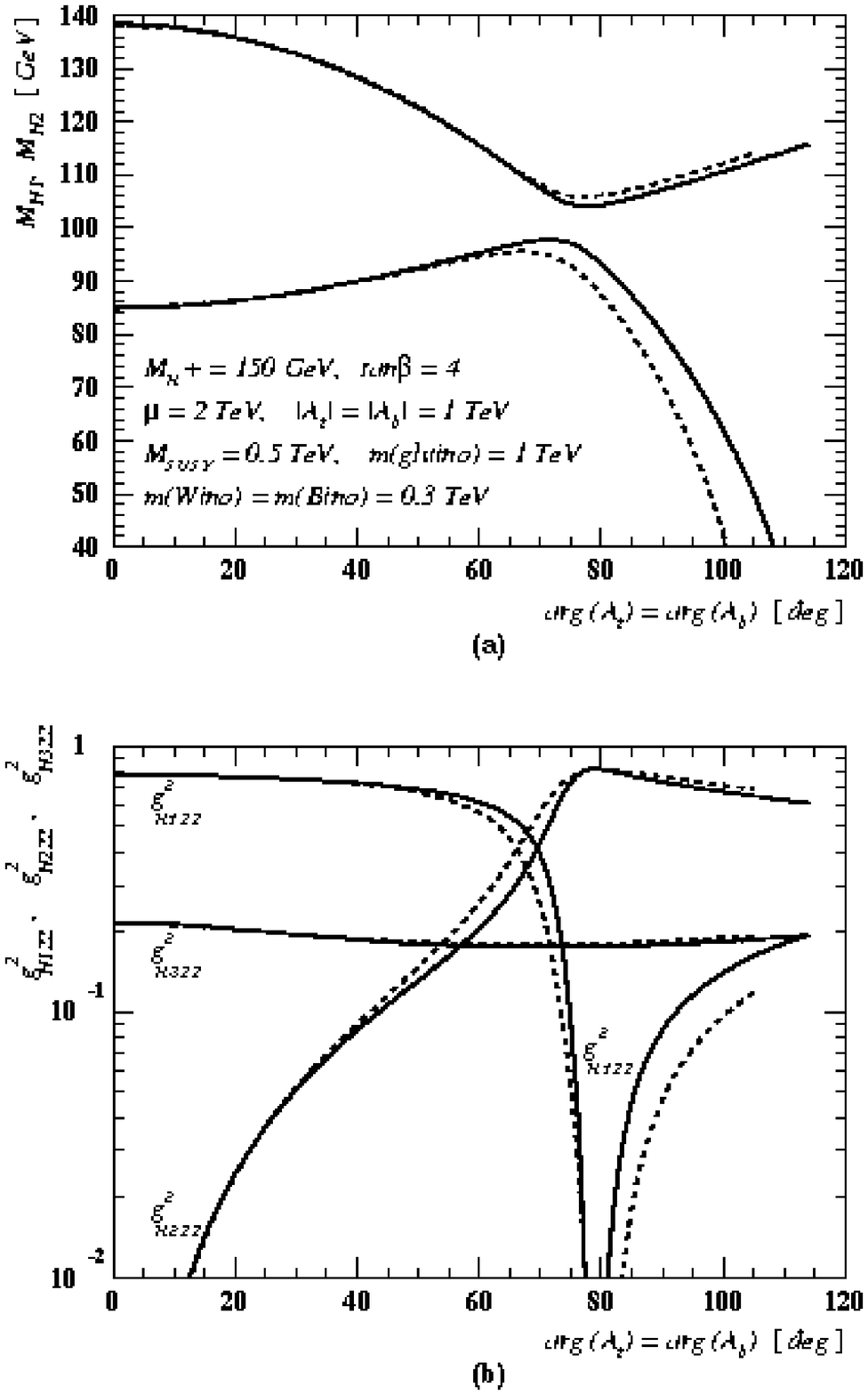,width=7cm}
\caption{\it Illustration how CP violation in the MSSM may suppress
the coupling of the lightest neutral Higgs boson to the $Z$~\cite{CEPW}.}
\label{fig:HZZ}
\end{figure}

\begin{figure}%Hbb
%\hglue4.5cm
\epsfig{figure=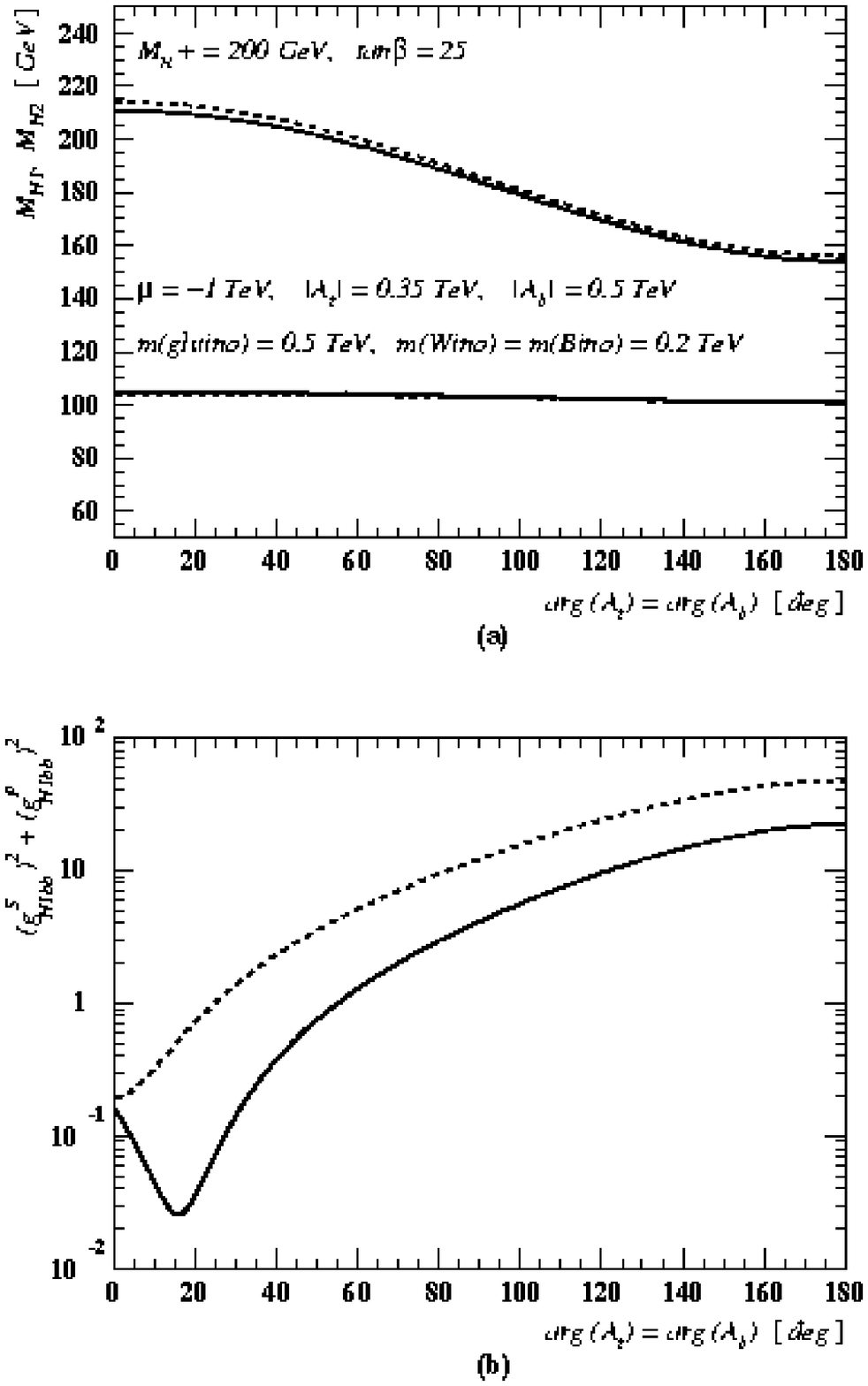,width=7cm}
\caption{\it Illustration how CP violation in the MSSM may suppress 
the coupling of the lightest neutral Higgs boson to
$\bar b b$~\cite{CEPW}.}
\label{fig:Hbb}
\end{figure}

The above analysis assumed that CP violation could be ignored. However, as
emphasized here by Kane~\cite{Kane}, CP violation in the soft
supersymmetry-breaking parameters may be important. Indeed, such CP
violation is essential in electroweak
baryogenesis~\cite{Losada}~\footnote{The popular
alternative of leptogenesis was discussed here by Ma~\cite{Ma}.}, which
also requires a first-order phase transition, and hence a relatively light
Higgs boson and stop squark, as seen in Fig.~\ref{fig:CQW}~\cite{CQW}! LEP
had been thought almost to exclude such a scenario because of its lower
limit on $m_h$. However, we have recently emphasized~\cite{CEPW} that the
LEP lower limit on $m_h$ may be greatly relaxed in the presence of CP
violation. For example, as seen in Fig.~\ref{fig:HZZ}, the h-H-A mixing
induced by CP violation may suppress the hZZ coupling, and the $h\bar bb$
coupling may also be suppressed, as seen in
Fig.~\ref{fig:Hbb}~\cite{CEPW}. We are currently re-evaluating the LEP
lower limit on $m_h$, taking these effects into account.

\section{Opportunities @ Future Accelerators}

\begin{figure}%LHCH
%\hglue4.5cm
\epsfig{figure=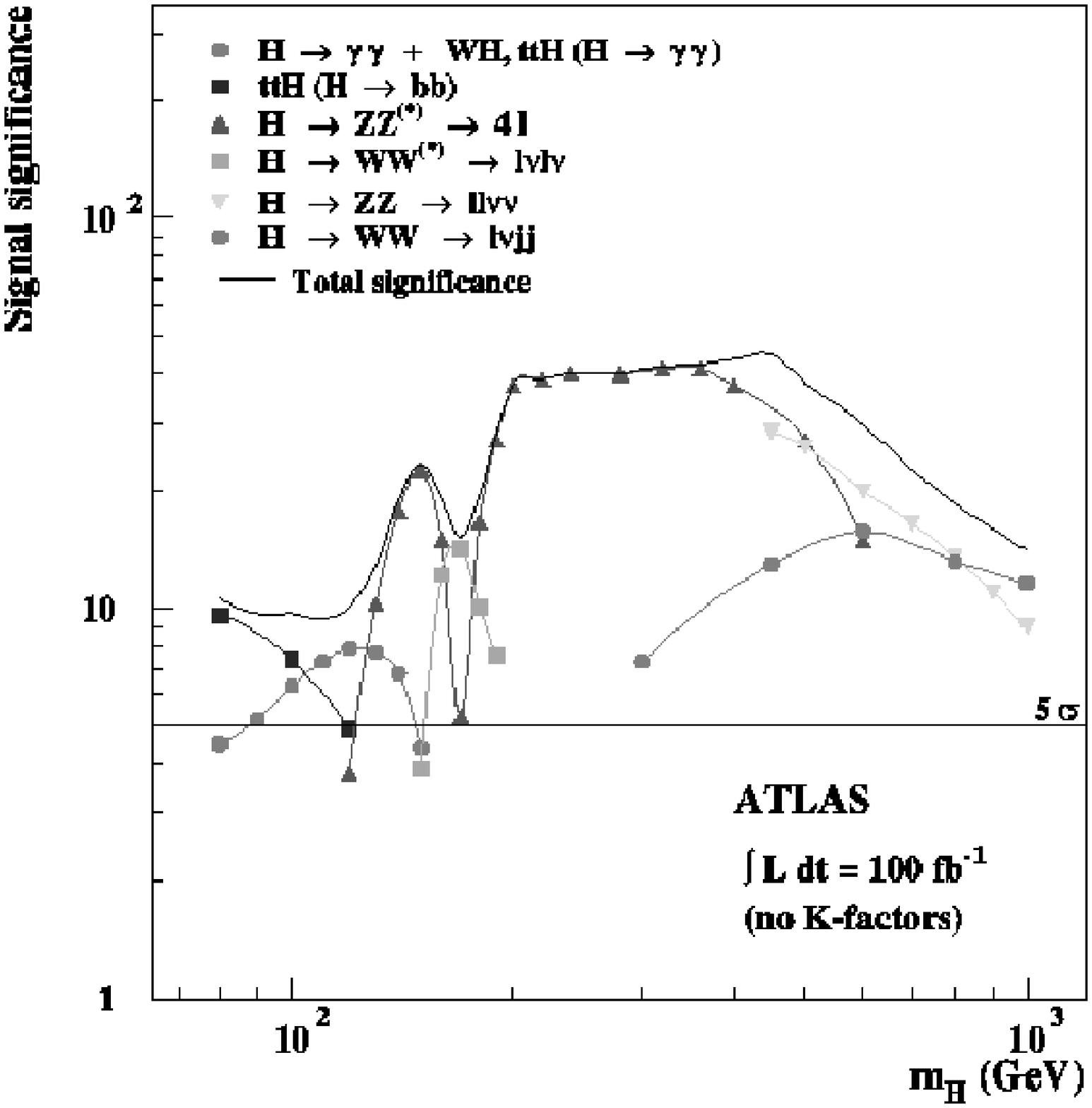,width=7cm}
\caption{\it Significance of the Standard Model Higgs signal
observable in the ATLAS experiment at the LHC~\cite{ATLAS}, for the
indicated value of the integrated luminosity.}
\label{fig:LHCH}
\end{figure}

As discussed here by Fernandez~\cite{Fernandez}, the LHC is under active
construction, and is scheduled to start operating in 2005. In this talk, I
concentrate on two selected LHC physics topics, namely the quest for the
Holy Higgs, and the search for supersymmetry. At low masses, the
$H\rightarrow\bar bb$ and $\gamma\gamma$ decay signatures look the most
promising, with $H\rightarrow 4 l^\pm$ over a large range of intermediate
masses, and $H\rightarrow W^+W^-\rightarrow l^+\nu l ^-\bar\nu, l^\pm \nu
jj$ and $H\rightarrow ZZ\rightarrow l^+l^- \bar\nu\nu$ interesting for
high masses. As seen in Fig.~\ref{fig:LHCH}~\cite{ATLAS}, there are no
holes in the mass coverage, a couple of decay modes can normally be
observed for any mass, and the Higgs mass can typically be measured with a
precision $10^{-3} \lappeq \Delta m_H/m_H \lappeq 10^{-2}$. The LHC will
also be able to discover supersymmetric Higgs bosons in two or more
channels, over all the supersymmetric parameter space.

\begin{figure}%LHCsusy
%\hglue4.5cm
\epsfig{figure=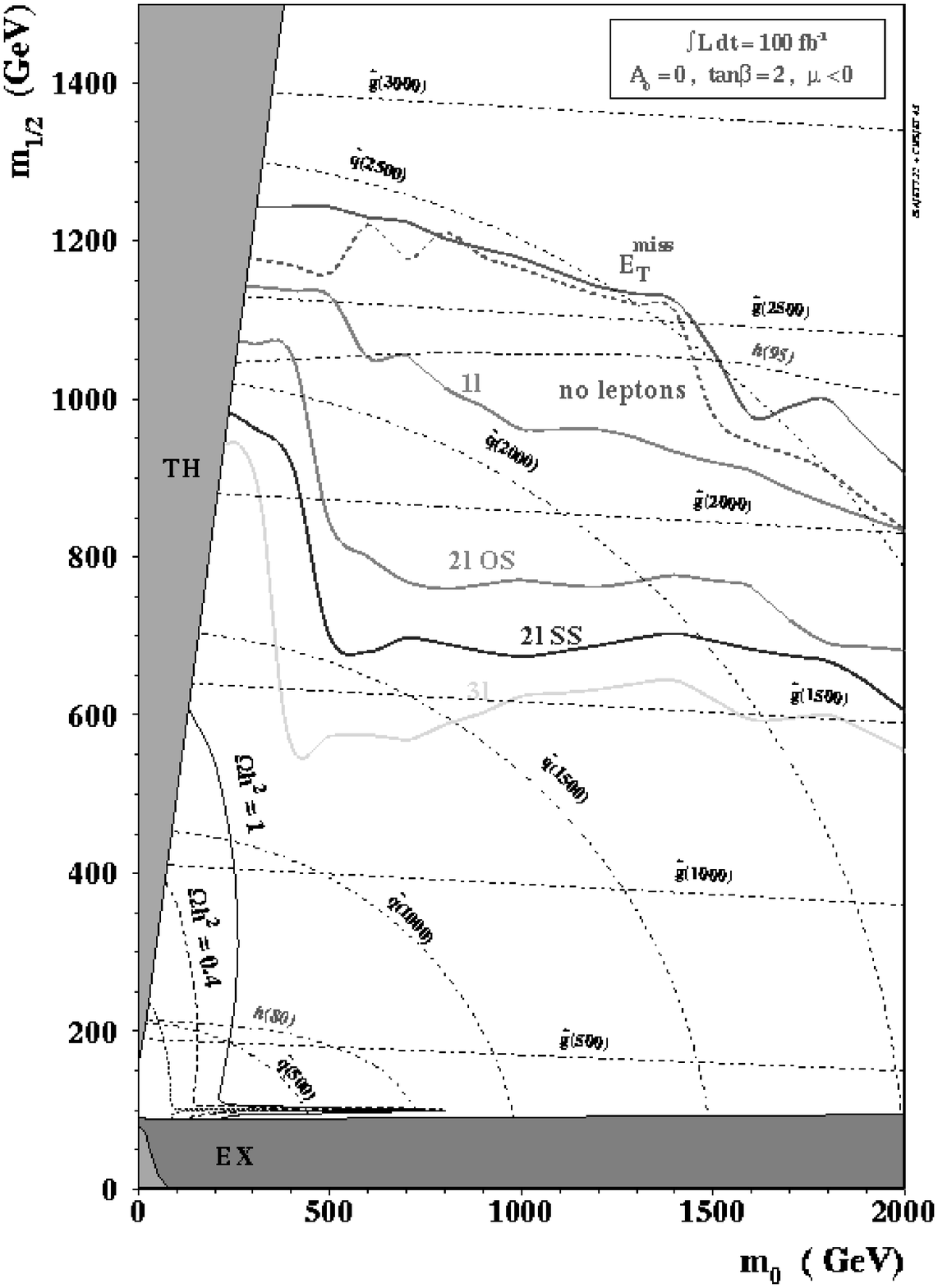,width=7cm}
\caption{\it Regions of the $m_0, m_{1/2}$ plane accessible to different
sparticle searches with the CMS experiment at the LHC~\cite{CMS}.}
\label{fig:LHCsusy}
\end{figure}

\begin{figure}%LHCcascade
%\hglue4.5cm
\epsfig{figure=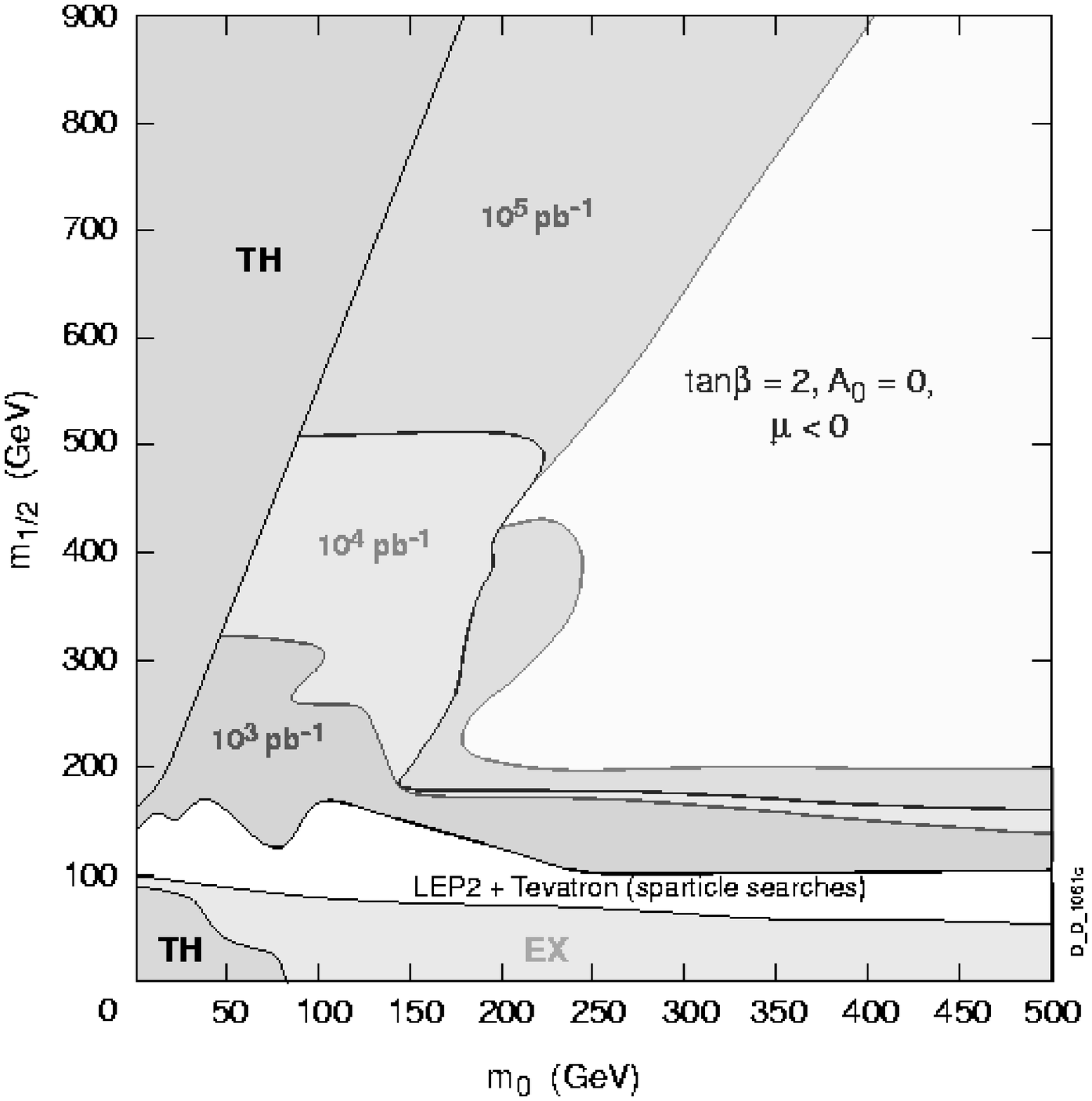,width=7cm}
\caption{\it Regions of the $m_0, m_{1/2}$ plane in which cascade
decays of sparticles are detectable with the CMS experiment at the
LHC~\cite{CMS}, for the indicated values of the integrated luminosity.}
\label{fig:LHCcascade}
\end{figure}

The LHC will produce principally strongly-interacting sparticles, squarks
$\tilde q$ and gluinos $\tilde g$, and they sould be detectable if they
weigh $\lappeq$ 2 TeV, as seen in Fig.~\ref{fig:LHCsusy}~\cite{CMS}. This
will enable the LHC to cover the parameter range allowed if the lightest
supersymmetric particle provides the cold dark matter. The $\tilde g$ and
$\tilde q$ often decay via complicated cascades, e.g., $\tilde
g\rightarrow \tilde b \bar b, \tilde b \rightarrow \chi^\prime b,
\chi^\prime \rightarrow \chi l^+l^-$, which may be reconstructed to
provide some detailed mass measurements, as indicated in
Fig.~\ref{fig:LHCcascade}~\cite{CMS}.

A plausible scenario for physics after the LHC is that the Higgs will have
been discovered, and one or two decay modes observed, and that several
sparticles will have been discovered, but that heavier Higgses, charginos
and sleptons may have escaped detection. 

\begin{figure}%Hdecays
%\hglue4.5cm
\epsfig{figure=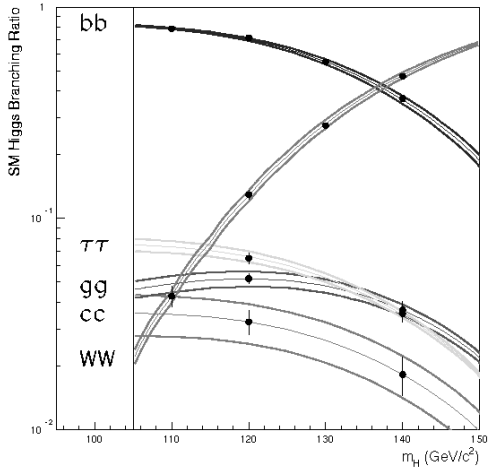,width=7cm}
\caption{\it The Standard Model Higgs branching ratios may be determined
very accurately at an $e^+ e^-$ linear collider~\cite{Battaglia}.}
\label{fig:Hdecays}
\end{figure}

These lacunae provide some of the motivation for $e^+e^-$ linear-collider
(LC)  physics. The very clean experimental environment, the egalitarian
production of new weakly-interacting particles and the prospective
availability of polarization make such a LC rather complementary to the
LHC~\cite{LC}. One of the big issues is what energy to choose for a
first-generation LC: we know there is the $\bar tt$ threshold at
$E_{cm}\simeq$ 350 GeV and we believe there should be a ZH threshold at
$E_{cm} = m_Z + m_H \lappeq$ 300 GeV. If one is above threshold, detailed
studies of many Higgs decay modes (as seen in
Fig.~\ref{fig:Hdecays}~\cite{Battaglia}), or measurements of sparticle
masses, become possible. However, we do not know what the supersymmetry
threshold might be (assuming there is one!). For this reason, I think it
is essential to retain as much flexibility as possible in the LC running
energy. 

\begin{figure}%LCsusy
%\hglue4.5cm
\epsfig{figure=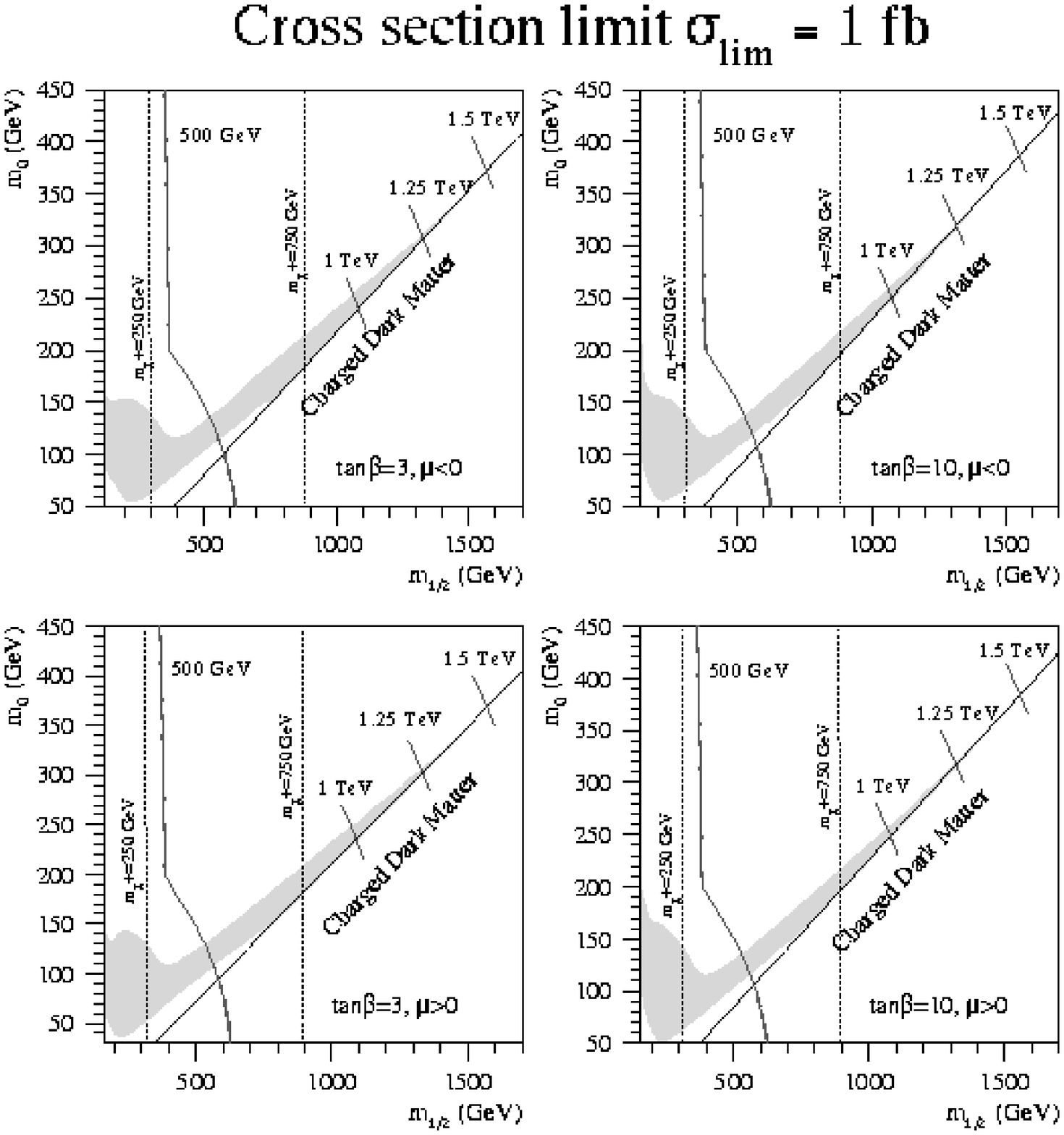,width=7cm}
\caption{\it Regions of the $m_{1/2}, m_0$ plane consistent with
cosmology (shaded) that may be explored with $e^+ e^-$ linear colliders
of the specified energies~\cite{EGO}, assuming universal soft
supersymmetry-breaking scalar masses.} 
\label{fig:LCsusy}
\end{figure}

In the meantime, the suggestion that the cold dark matter might consist of
supersymmetric particles can be used~\cite{EGO} to guess the likelihood
that a LC of given energy might find supersymmetry. As seen in
Fig.~\ref{fig:LCsusy}, we find that all the dark matter parameter space
can be explored if $E_{cm} \geq$ 1.25~TeV, and about 90 \% if $E_{cm}$ = 1
TeV, but that an LC with $E_{cm}$ = 0.5 TeV would only cover about 60 \%
of the dark matter parameter region.

My opinion is that physics will demand an LC in the TeV energy range, and
I hope that the world can converge on a (single) project in this energy
range. In the rest of this talk, I assume that an LC with $E_{cm} \sim$ 1
TeV will be built somewhere in the world, and ask what other accelerator
projects might be interesting~\cite{EKR}. 

One suggestion is a future larger hadron collider with 100 TeV $\lappeq
E_{cm} \lappeq$ 200 TeV, that could explore the 10 TeV mass region for the
first time, if its luminosity rises to $10^{35}$ cm$^{-2}$ s$^{-1}$ or so.
Such a machine is probably technically feasible, but it would be enormous,
with a circumference of 100 to 500 km. The principal challenge will be
reducing the unit cost by an order of magnitude compared to the LHC. At
present, we cannot formulate the physics questions for such a machine with
great clarity. 

\begin{figure}%Marco
%\hglue4.5cm
\epsfig{figure=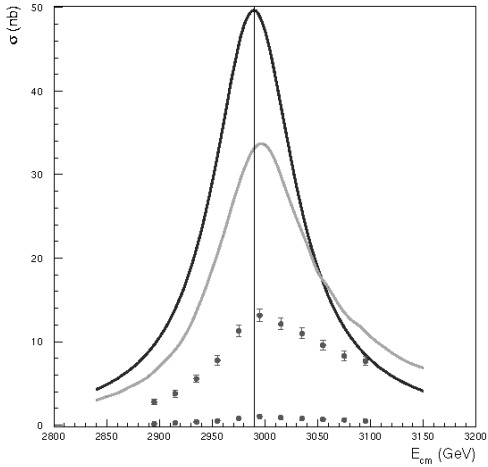,width=7cm}
\caption{\it First impression of a possible $Z'$ scan at CLIC, showing
how beamstrahlung reduces the cross section below the value expected if
only initial-state radiation is included~\cite{CLICZ}.}
\label{fig:Marco}
\end{figure}

Another possibility is a higher-energy LC with $E_{cm} \gappeq$ 2 TeV,
capable, e.g., of making precise and complete studies of sparticle
spectroscopy, or of any other electroweakly-interacting sparticles
weighing $\lappeq$ 1 TeV.  CERN is developing a potential technology for
such an LC, called CLIC~\cite{CLIC}, in which an intense low-energy drive
beam is used to provide RF to accelerate a more energetic but less intense
colliding beam.  Accelerating gradients in the range 100 to 200 MeV
$m^{-1}$ appear possible, enabling an LC with $E_{cm} \lappeq$ 5 TeV to be
accommodated in a tunnel $\sim$ 35 km long. The first physics study for
such a machine was made at La Thuile in 1987~\cite{LaThuile}, and a new
physics study has now been initiated~\cite{CLICphysics}. 
Fig.~\ref{fig:Marco} is a first result from this new study, showing what a
$Z'$ resonance might look like at CLIC~\cite{CLICZ}. 

\begin{figure}%musr
%\hglue4.5cm
\epsfig{figure=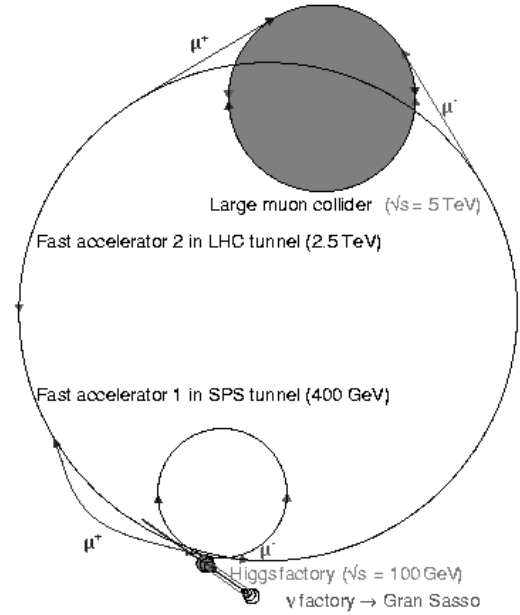,width=7cm}
\caption{\it Artist's impression how a muon storage ring complex
could be accommodated near CERN, including a $\nu$ factory, a Higgs
factory and a high-energy muon collider~\cite{ABE}.}
\label{fig:musr}
\end{figure}

A third type of accelerator currently attracting much interest is a
complex of muon storage rings, as illustrated in Fig.~\ref{fig:musr}. 
These could be developed in three steps~\cite{ABE}: first a neutrino
factory in which muons are simply allowed to decay without colliding,
secondly a Higgs factory colliding $\mu^+\mu^-$ at $E_{cm} \sim m_H$, and
thirdly a high-energy collider which might be compared with CLIC as a
device to probe the high-energy frontier. The chief advantages of neutrino
beams from $\mu$ decays, as opposed to conventional beams from hadron
decays, are their precisely calculable fluxes and spectra, and the facts
that equal numbers of $\nu_\mu$ and $\bar\nu_e$ (or $\bar\nu_\mu$ and
$\nu_e$)  are produced. A $\mu^+\mu^-$ collider used as a Higgs factory
can measure very precisely the mass and width of any Higgs boson with mass
around 100 GeV, distinguishing between the Standard Model and a
superymmetric extension, and measurements of supersymmetric Higgs peaks
could provide a unique window on CP violation. A $\mu^+\mu^-$ collider at
the high-energy frontier has advantages over an $e^+e^-$ LC with similar
energy, conferred by the more precise energy calibration and reduced
energy spread, but the ultimate energy is limited by the neutrino
radiation hazard. However, although it is very attractive, many technical
problems need to be solved before the feasibility of such a muon storage
rign complex can be established.

\begin{figure}%clip
%\hglue4.5cm
\epsfig{figure=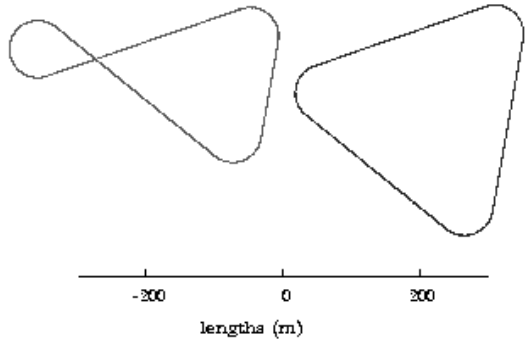,width=7cm}
\caption{\it Vertical sections through possible designs of muon storage
rings for $\nu$ factories: the re-entrant design is advantageous for
retaining muon beam polarization~\cite{bentclip}.}
\label{fig:clip}
\end{figure}

The basic concept for a neutrino factory involves a proton driver with
beam power 1 to 20 MW, provided by either a linac or a rapid-cycling
synchrotron. This is used to produce pions, which decay into muons, of
which about 0.1/proton are cooled, accelerated to 10 to 50 GeV, and stored
in a ring. This need not be circular, and may look more like a bent
paper-clip, as seen in Fig.~\ref{fig:clip}~\cite{bentclip}, with several
straight sections sending $\sim (10^{20}$ to $10^{21}) \bar\nu_\mu, \nu_e$
per year each towards detectors at different distances. 

\begin{figure}%mue
%\hglue4.5cm
\epsfig{figure=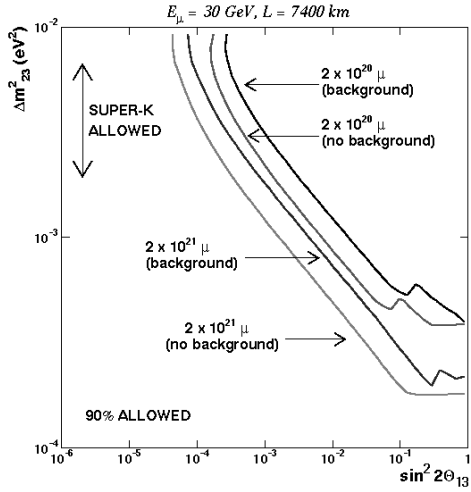,width=7cm}
\caption{\it Sensitivities to $\nu_\mu \rightarrow \nu_e$
oscillations with a $\nu$ factory~\cite{BCR}.}
\label{fig:mue}
\end{figure}

\begin{figure}%CP
%\hglue4.5cm
\epsfig{figure=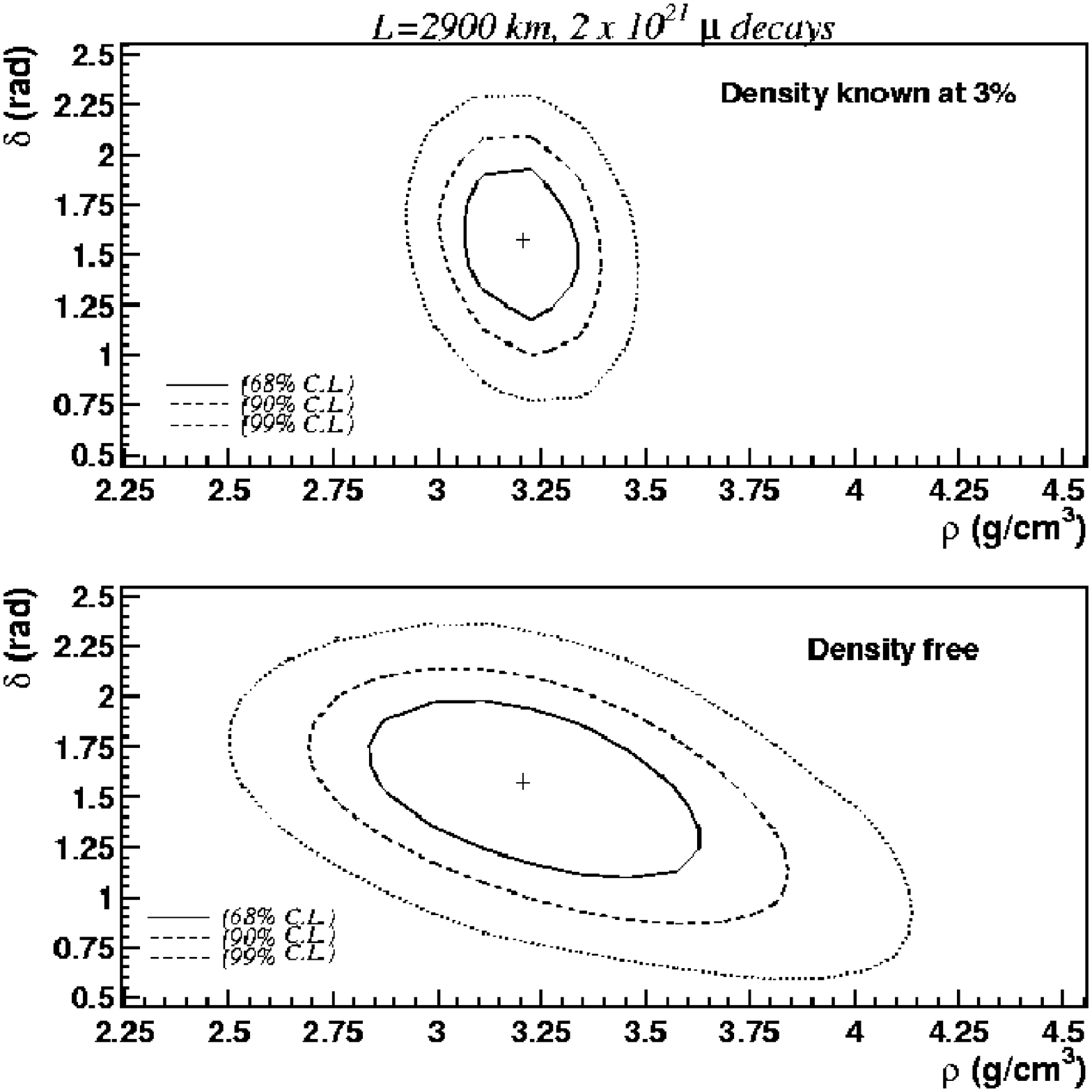,width=7cm}
\caption{\it Sensitivity to CP violation in $\nu$
oscillations with a $\nu$ factory, either with or without
knowledge of the Earth's matter density~\cite{BCR}.}
\label{fig:CP}
\end{figure}

\begin{figure}%web
%\hglue4.5cm
\epsfig{figure=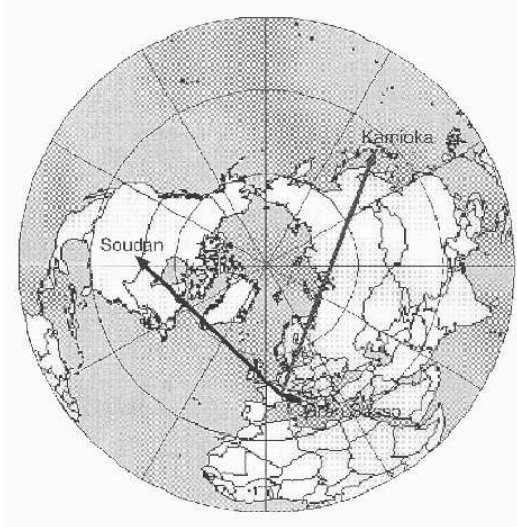,width=7cm}
\caption{\it Artist's impression of a possible `World-Wide Neutrino Web'
of
long-baseline $\nu$ beams aimed at experiments in different regions of
the world.} 
\label{fig:web}
\end{figure}

In long-baseline neutrino experiments with a neutrino factory, the
sensitivities to mixing angles and $\Delta m^2$ depend on $E_\mu$ and the
distance $L$~\cite{DGH}.
As we see in Fig.~\ref{fig:mue}, a $\nu_\mu\rightarrow\nu_e$
appearance experiment would be much more sensitive than the present
Super-Kamiokande upper limit~\cite{BCR}, or what may be achieved with
MINOS. Moreover, as seen in Fig.~\ref{fig:CP}, with sufficiently many
$\mu$ decays one may be sensitive to CP violation and matter effects
(which depend on the sign of $\Delta m^2$) in neutrino oscillations. For
this, a detector at a distance of 2000 to 5000 km would be particularly
advantageous. Ultimately, one could imagine a ``World-Wide Neutrino Web"
consisting of a $\nu$ factory in one region of the world feeding detectors
in the same and other regions, as illustrated in Fig.~\ref{fig:web}.

\begin{figure}%Hpeak
%\hglue4.5cm
\epsfig{figure=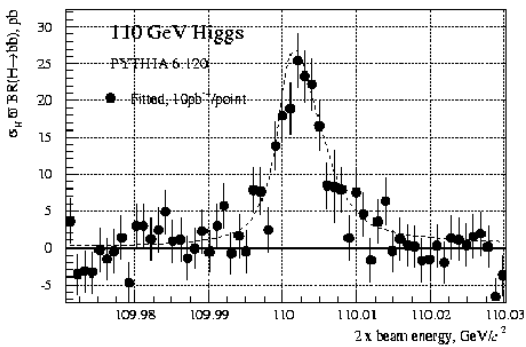,width=7cm}
\caption{\it The possible line-shape of the Standard Model Higgs peak at a
$\mu^+ \mu^-$ collider operated as a Higgs factory~\cite{ABE}.}
\label{fig:Hpeak}
\end{figure}

\begin{figure}%twinpeak
%\hglue4.5cm
\epsfig{figure=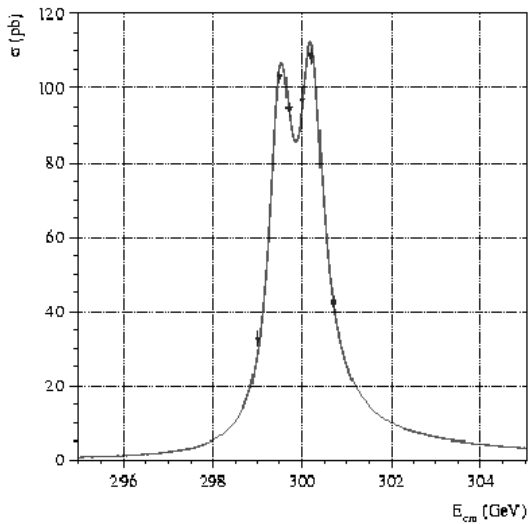,width=7cm}
\caption{\it The possible line-shape of the twin MSSM Higgs peaks at
a $\mu^+ \mu^-$ collider operated as a second-generation Higgs
factory~\cite{ABE}.}
\label{fig:twinpeak}
\end{figure}

Turning now to a Higgs factory, in the absence of a beam energy spread,
the line shape (see Fig.~\ref{fig:Hpeak}) should be 
\beq 
\sigma_H (s) =
{4\pi \Gamma(H\rightarrow\mu^+\mu^-) \Gamma(H\rightarrow X)\over
(s-m^2_H)^2 + m^2_H \Gamma^2_H} 
\label{twenty} 
\eeq 
It seems that it might
be possible to reduce the beam energy spread to $\sim$ 0.01 \% or 10 MeV,
comparable to the natural width of 3 MeV for a Standard Model Higgs
weighing about 100 GeV. Calibrating the beam energy via the $\mu^\pm$
polarization, it should then be possible to measure $m_H$ with a precision
of $\pm$ 0.1 MeV, and the width to within 0.5 MeV, sufficient to
distinguish a Standard Model Higgs boson from the lightest supersymmetric
Higgs boson, over a large range of parameter space~\cite{ABE}. In the
supersymmetric case, a second-generation Higgs factory able to explore the
``Twin Peaks" of the $H$ and $A$ shown in Fig.~\ref{fig:twinpeak} might be
even more interesting, providing tests of CP symmetry analogous to those
in the $K^0-\bar K^0$ system~\cite{Pilaftsis}.

At the high-energy frontier, $\mu^+\mu^-$ colliders would benefit from the
absence of beamstrahlung and reduced initial-state radiation, as compared
to an $e^+e^-$ LC such as CLIC. However, the latter offers controllable
beam polarization, $e\gamma$ and $\gamma\gamma$ colliders ``for free", and
avoids the problems presented by $\mu$ decays.  Moreover, an $e^+e^-$ LC
demonstrator, namely the SLC, has been built, whereas many of the
technologies needed for a $\mu^+\mu^-$ collider are at best glints in the
eye, at present. 

\section{Towards a Theory of Everything?}

The job description of a Theory of Everything (TOE) is to unify all the
fundamental interactions, including gravity, and to solve all the problems
that arise in attempts to quantize gravitation. A possible solution is to
replace point-like elementary particles by extended objects, and the first
incarnation of this idea used one-dimensional closed loops of string. It
was soon realized that this scenario requires extra space-time dimensions
10 = 4 + 6 in the supersymmetric case, and/or extra interactions. The
current reincarnation of this idea as $M$ theory includes other extended
objects such as two-dimensional membranes, solids, etc.~\cite{Russo}. 

The key question is how to test these ideas. A popular suggestion is that
the 6 surplus space dimensions are compactified on a Calabi-Yau manifold,
but which one? 473~800~776 are known~\cite{KS}! Recently we have embarked
on a systematic study of Calabi-Yau (CY) spaces, constructed as zeroes of
polynomials in weighted projective spaces, a technique which enables one
to explore some of their internal properties and focus on these with
desirable features~\cite{AENV}. Our harvest so far comprises 182~737
CY spaces, of which 211 have 3 generations
and K3 fibrations as desired in some approaches to $M$ theory~\cite{AENV}. 

%\begin{figure}%ke
%\hglue4.5cm
%\epsfig{figure=skarkel2.eps,width=7cm}
%\caption{\it Scatter plot of the topological numbers of the
%first harvest of Calabi-Yau manifolds constructed explicitly
%in~\cite{AENV}.}
%\label{fig:Skarke}
%\end{figure}

In the face of this ambiguity, how can one speak of phenomenological
predictions from string theory? In fact, it has told us correctly that
there cannot be more than 10 dimensions, that the gauge group cannot be
very large, that matter representations cannot be very big, and that the
top quark should not weigh more than about 190 GeV. It has also provided a
first-principles estimate of the unification scale, $M_U \sim$ few $\times
10^{17}$ GeV~\cite{mstring}, which is not so far from the phenomenological
bottom-up estimate of (1 or 2) $\times 10^{16}$ GeV. 

Moreover, we now understand that all the different string theories are
related by dualities in the general framework known as $M$
theory~\cite{Mtheory}. The question then becomes, in which part of its
parameter space do we live? The GUT mass-scale calculation suggests that
the string coupling may be strong~\cite{HW}, corresponding to one large
dimension: $L \gg 1/M_{GUT} \simeq 1/10^{16}$ GeV $\gg 1/m_P \equiv l_P$. 

Adventurous souls have then gone on to propose that one or more ``small" 
dimensions might actually be rather large, perhaps $L \sim 1$ TeV$^{-1}$
or even $\sim$ 1 mm~\cite{golow}? In such models, there may be observable
modifications of Newton's law: $G_N/r \rightarrow G_N(1/r)(L/r)^{\Delta
D}$ at short distances, as well as possible new accelerator signatures.
This suggestion offers plenty of phenomenological fun, but why should the
scale of gravity sink so low? Are there any advantages for the hierarchy
problem? So far, I have seen it reformulated, but not yet solved. 

Before closing, I would like to mention a couple of radical possibilities
for string phenomenology. As we heard here, surprisingly many
ultra-high-energy (UHE)  cosmic rays have been observed above $E\sim
5\times 10^{19}$ eV, more than expected above the GZK cutoff due to the
reaction $p + \gamma_{CMB} \rightarrow \Delta^+$~\cite{Zepeda}. Unless one
modifies relativistic kinematics (see later) these UHE cosmic rays should
have originated nearby, at distances $d \lappeq$ 100 Mpc for $E \sim
10^{20}$ eV, but no discrete sources have yet been confirmed.

\begin{figure}%BS
%\hglue4.5cm
\epsfig{figure=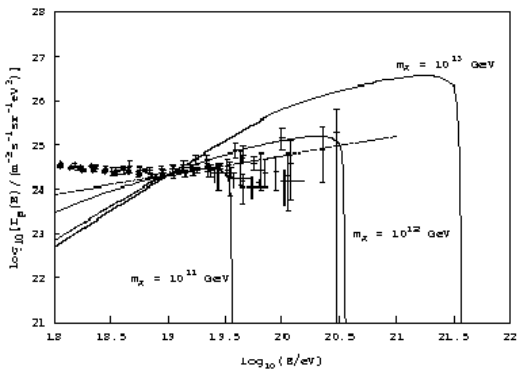,width=7cm}
\caption{\it Comparison between data on ultra-high-energy cosmic rays
around the GZK cutoff with a simulation of crypton decays~\cite{BS}.}
\label{fig:BS}
\end{figure}

Might they originate from the decays of supermassive dark matter
particles? It has recently been realized that such particles weighing
10$^{10}$ GeV or more might have been produced by non-thermal mechanisms
early in the history of the Universe~\cite{CKR}. Possible candidates for
these unstable heavy relics can be found in string theory, particularly as
bound states in the hidden sector, which we have termed
cryptons~\cite{cryptons}. They could well have masses $\sim 10^{13}$ GeV
and be metastable, decaying via higher-dimensional operators into multiple
leptons and quarks. Simulations indicate that they could well produce the
observed UHE tail of the cosmic-ray spectrum, as seen in
Fig.~\ref{fig:BS}~\cite{BS}. The Auger project should be able to tell us
whether this mechanism is tenable~\cite{Zepeda}.

\begin{figure}%astro
%\hglue4.5cm
\epsfig{figure=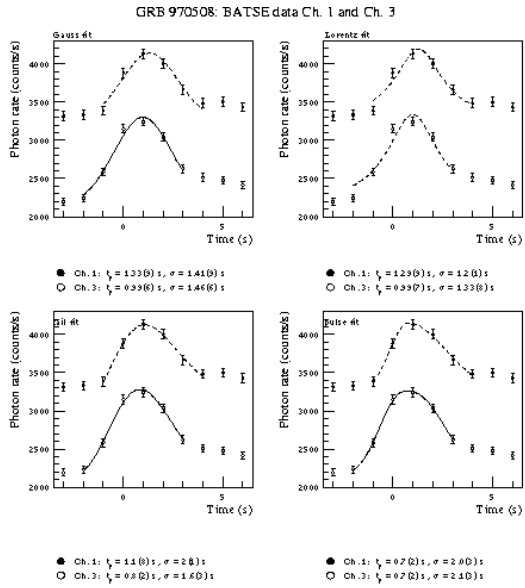,width=7cm}
\caption{\it Fits to BATSE data on GRB 970508 in different energy
channels~\cite{EFMMN}. Comparisons between the arrival times of the
peaks at different energies are used to constrain any possible
energy dependence of the velocity of light.}
\label{fig:EFMMN}
\end{figure}

Even more speculative is the suggestion of quantum-gravitational
phenomenology.  Might the space-time foam of quantum-gravitational
fluctuations in the fabric of space induce quantum decoherence and/or CPT
violation at the microscopic scale~\cite{EMN}?  Here the most sensitive
probe may be the $K^0-\bar K^0$ system~\cite{EHNS}. Might the velocity of
light~\cite{AEMN} (or a neutrino~\cite{EMNV}) depend on its energy,
because of recoil effects on the space-time vacuum? Here the most
sensitive direct tests may be provided by distant, energetic sources with
short time-scales, such as gamma-ray bursters (GRBs)~\cite{AEMNS}, active
galactic nuclei~\cite{AGNs} and pulsars~\cite{Kaaret}, and some such
models are also constrained by the kinematics of UHE cosmic
rays~\cite{Mestres}. Fig.~\ref{fig:EFMMN} shows fits to BATSE data
on GRB~970508 in different energy channels. A regression analysis of fits
to GRBs with measured redshifts has been used to constrain any possible
energy dependence of the velocity of light: $\delta c / c \le E/M: M >
10^{15}$~GeV~\cite{EFMMN}.

\section{Final Comments}

The history of physics reveals many ways in which it may advance, being
driven either by pure theoretical thought or by experimental
breakthroughs. Pure theoretical speculation must in any case be tempered
by experimental reality: we can never forget that in physics, as any other
science, experiment is the ultimate arbiter. Particle physics is currently
fortunate. On the one hand, experiments at LEP and elsewhere have shown
that the Standard Model is a solid rock on which to build. On the other
hand, experiments on neutrinos strongly indicate oscillations, and hence
physics beyond the Standard Model. There are exciting new experimental
programmes underway to explore the flavour problem, to pin down models of
neutrino oscillations, and to explore the TeV energy range. 

Beyond our daily concerns, we have the great unsolved problem of
twentieth-century physics left to stimulate us: reconcile gravity and
quantum mechanics. Great theoretical advances towards this goal have been
made during recent years, but we do not know how far we are from this
goal. In particular, we have not yet defined a clear experimental test
that will confirm or refute string theory. Finding it is our key
phenomenological challenge.

\end{document}